%Paper: hep-th/9506201
%From: davbmath@durras.anu.edu.au
%Date: Fri, 30 Jun 95 10:32:59 EST

% ----------------------------------------------------------------------------
%
% Plain TeX.
%
% 1. For correct cross-references, run twice (as with LaTeX).
%
% 2. The FIGURES for this paper are at the end of this file.
%    They must split off and saved with names fig1.eps, fig2.eps, fig3.eps
%    The standard macros package epsf.tex is also required.
%
% ----------------------------------------------------------------------------
%
\input epsf.tex
%
% ----------------------------------------------------------------------------
%	This file contains definitions for Euler Fraktur and Black Board Bold fonts.
% ----------------------------------------------------------------------------
%
\catcode`\@=11

\font\tenmsy=msbm10
\font\sevenmsy=msbm10 scaled 700
\font\fivemsy=msbm10 scaled 500
\newfam\msyfam
\textfont\msyfam=\tenmsy
\scriptfont\msyfam=\sevenmsy
\scriptscriptfont\msyfam=\fivemsy

\def\hexnumber@#1{\ifcase#1 0\or1\or2\or3\or4\or5\or6\or7\or8\or9\or
	A\or B\or C\or D\or E\or F\fi }

\font\teneuf=eufm10
\font\seveneuf=eufm7
\font\fiveeuf=eufm5
\newfam\euffam
\textfont\euffam=\teneuf
\scriptfont\euffam=\seveneuf
\scriptscriptfont\euffam=\fiveeuf
\def\goth{\let\next\relax\ifmmode\let\next\goth@\else
\message{Use goth only in math mode}\fi\next}
\def\goth@#1{{\goth@@{#1}}}
\def\goth@@#1{\fam\euffam#1}

\def\Bbb{\ifmmode\let\next\Bbb@\else
\def\next{\message{Use Bbb only in math mode}}\fi\next}
\def\Bbb@#1{{\Bbb@@{#1}}}
\def\Bbb@@#1{\fam\msyfam#1}

\catcode`\@=12

\message{Cross-reference macros, B. Davies, version 5 May 1993.}

\catcode`@=11

% look for xrf file from previous run

\newif\if@xrf\@xrffalse   % this becomes true once the xrf file is set up
\def\l@bel #1 #2 #3>>{\expandafter\gdef\csname @@#1#2\endcsname{#3}}
\immediate\newread\xrffile
\def\xrf@n#1#2{\expandafter\expandafter\expandafter
\csname immediate\endcsname\csname #1\endcsname\xrffile#2}
\def\xrf@@n{\if@xrf\relax\else%
  \expandafter\xrf@n{openin}{ = \jobname.xrf}\relax%
  \ifeof\xrffile%
    \message{ no file \jobname.xrf - run again for correct forward references
}%
  \else%
    \expandafter\xrf@n{closein}{}\relax%
    \setbox0=\hbox{\catcode`@=11 \input\jobname.xrf \catcode`@=12}%
  \fi\global\@xrftrue%
  \expandafter\expandafter\csname immediate\endcsname%
  \csname  newwrite\endcsname\xrffile%
  \expandafter\xrf@n{openout}{ = \jobname.xrf}\relax\fi}

% general macros which define \<string> and \ref<string> from \order{string}
% eg. \order{thm} defines \thm{label} and \refthm{label}

\newcount\t@g

\def\order#1{%
  \expandafter\expandafter\csname newcount\endcsname
  \csname t@ghd#1\endcsname\csname t@ghd#1\endcsname=0

  \expandafter\def\csname #1\endcsname##1{\xrf@@n\csname n@#1\endcsname##1:>}

  \expandafter\def\csname n@#1\endcsname##1:##2>%
    {\def\n@xt{##1}\ifx\n@xt\empty%
     \expandafter\csname n@@#1\endcsname##1:##2:>
     \else\def\n@xt{##2}\ifx\n@xt\empty%
     \expandafter\csname n@@#1\endcsname\unp@ck##1 >:##2:>\else%
     \expandafter\csname n@@#1\endcsname\unp@ck##1 >:##2>\fi\fi}

  \expandafter\def\csname n@@#1\endcsname##1:##2:>%
    {\edef\t@g{\csname t@g#1\endcsname}\edef\t@@ghd{\csname t@ghd#1\endcsname}%
     \ifnum\t@@ghd=\t@ghd\else\global\t@@ghd=\number\t@ghd\global\t@g=0\fi%
     \ifunc@lled{@#1}{##1}\global\advance\t@g by 1%
       {\def\n@xt{##1}\ifx\n@xt\empty%
       \else\writ@new{#1}{##1}{\pret@g\t@ghead\number\t@g}\expandafter%
       \xdef\csname @#1##1\endcsname{\pret@g\t@ghead\number\t@g}\fi}%
       {\pret@g\t@ghead\number\t@g}%
     \else\def\n@xt{##1}%
       \w@rnmess#1,\n@xt>\csname @#1##1\endcsname%
     \fi##2}\ord@r{#1}}

\def\ord@r#1{%
  \expandafter\expandafter\csname newcount\endcsname
  \csname t@g#1\endcsname\csname t@g#1\endcsname=0

  \expandafter\def\csname ref#1\endcsname##1{%
     \expandafter\each@rg\csname #1c@te\endcsname{##1}}

  \expandafter\def\csname #1c@te\endcsname##1:##2,%
    {\def\n@xt{##2}\ifx\n@xt\empty%
     \csname #1cit@\endcsname##1:##2:,\else%
       \csname #1cit@\endcsname##1:##2,\fi}

  \expandafter\def\csname #1cit@\endcsname##1:##2:,%
    {\def\n@xt{\unp@ck##1 >}\ifunc@lled{@#1}{\n@xt}%
      {\expandafter\ifx\csname @@#1\n@xt\endcsname\relax%
       \und@fmess#1,\n@xt>>>\n@xt<<%
       \else\csname @@#1\n@xt\endcsname##2\fi}%
     \else\csname @#1\n@xt\endcsname##2%
     \fi}}

% same as \order{string} except that there are no prefixes

\def\sporder#1{%

  \expandafter\def\csname #1\endcsname##1{\xrf@@n\csname n@#1\endcsname##1:>}

  \expandafter\def\csname n@#1\endcsname##1:##2>%
    {\def\n@xt{##1}\ifx\n@xt\empty%
     \expandafter\csname n@@#1\endcsname##1:##2:>
     \else\def\n@xt{##2}\ifx\n@xt\empty%
     \expandafter\csname n@@#1\endcsname\unp@ck##1 >:##2:>\else%
     \expandafter\csname n@@#1\endcsname\unp@ck##1 >:##2>\fi\fi}

  \expandafter\def\csname n@@#1\endcsname##1:##2:>%
    {\edef\t@g{\csname t@g#1\endcsname}%
     \ifunc@lled{@#1}{##1}\global\advance\t@g by 1%
       {\def\n@xt{##1}\ifx\n@xt\empty%
       \else\writ@new{#1}{##1}{\number\t@g}\expandafter%
       \xdef\csname @#1##1\endcsname{\number\t@g}\fi}{\number\t@g}%
     \else\def\n@xt{##1}\w@rnmess#1,\n@xt>\csname @#1##1\endcsname%
     \fi##2}\ord@r{#1}}

% for dealing with list of arguments separated by commas and
% dashes, and applying a specified control sequence to each
% so we can have \ref{a,b}, etc

\def\each@rg#1#2{{\let\thecsname=#1\expandafter\first@rg#2,\end,}}
\def\first@rg#1,{\callr@nge{#1}\apply@rg}
\def\apply@rg#1,{\ifx\end#1\let\n@xt=\relax%
\else,\callr@nge{#1}\let\n@xt=\apply@rg\fi\n@xt}

\def\callr@nge#1{\calldor@nge#1-\end-}
\def\callr@ngeat#1\end-{#1}
\def\calldor@nge#1-#2-{\ifx\end#2\thecsname#1:,%
  \else\thecsname#1:,\hbox{\rm--}\thecsname#2:,\callr@ngeat\fi}

% for turning any labels of the form \alpha into @alpha

\def\unp@ck#1 #2>{\unp@@k#1@> @>>}
\def\unp@@k#1 #2>>{\ifx#2@\@np@@k#1\else\@np@@k#1@> \unp@@k#2>>\fi}
\def\@np@@k#1#2#3>{\ifx#2@\@@np@@k#1>\else\@@np@@k#1>\@np@@k#2#3>\fi}
\def\@@np@@k#1>{\ifcat#1\alpha\expandafter\@@np@@@k\string#1>\else#1\fi}
\def\@@np@@@k#1#2>{@#2}

% for writing labels to jobname.xrf

\def\writ@new#1#2#3{\xrf@@n\immediate\write\xrffile
  {\noexpand\l@bel #1 #2 {#3}>>}}

% for checking if labels are undefined

\def\ifunc@lled#1#2{\expandafter\ifx\csname #1#2\endcsname\relax}
\def\und@fmess#1#2,#3>{\ifx#1@%
  \message{ ** error - eqn label >>#3<< undefined - run again ** }\else
  \message{ ** error - #1#2 label >>#3<< undefined - run again ** }\fi}
\def\w@rnmess#1#2,#3>{\ifx#1@%
  \message{ Warning - duplicate eqn label >>#3<< }\else
  \message{ Warning - duplicate #1#2 label >>#3<< }\fi}

% to precede equation numbers by a taghead

\def\t@ghead{}
\newcount\t@ghd\t@ghd=0
\def\taghe@d#1{\gdef\t@ghead{#1}\global\advance\t@ghd by 1}

% define equation labels with an @qn

\order{@qn}

% now define \eqno(label), \leqno(label), \ref(label)
% also redefine \eqalignno & \leqalignno so that &() is picked up

\let\eqno@@=\eqno
\def\eqno(#1){\xrf@@n\eqno@@\hbox{{\rm(}$\@qn{#1}${\rm)}}}

\let\leqno@@=\leqno
\def\leqno(#1){\xrf@@n\leqno@@\hbox{{\rm(}$\@qn{#1}${\rm)}}}

\def\ref#1{\xrf@@n{{\rm(}$\ref@qn{#1}${\rm)}}}

% now define \eq(label) for IOPP preprint.sty

\def\eq(#1){\xrf@@n\hfill\llap{\hbox{{\rm(}$\@qn{#1}${\rm)}}}}

% the following only differs from plain TeX by picking up equation labels
% \c@ntering is the same as \centering but does not conflict with LaTeX

\expandafter\csname newskip\endcsname\c@ntering
\c@ntering=0pt plus 1000pt minus 1000pt
\def\eqalignno#1{\xrf@@n\displ@y \tabskip=\c@ntering
  \halign to\displaywidth{\hfil$\displaystyle{##}$\tabskip=0pt
   &$\displaystyle{{}##}$\hfil\tabskip=\c@ntering
   &\llap{$\eqaln@##$}\tabskip=0pt\crcr
   #1\crcr}}
\def\leqalignno#1{\xrf@@n\displ@y \tabskip=\c@ntering
  \halign to\displaywidth{\hfil$\displaystyle{##}$\tabskip=0pt
   &$\displaystyle{{}##}$\hfil\tabskip=\c@ntering
    &\kern-\displaywidth\rlap{$\eqaln@##$}\tabskip\displaywidth\crcr
   #1\crcr}}
\def\eqaln@#1#2{\relax\ifcat#1(\expandafter\eqno@\else\fi#1#2}
\def\eqno@(#1){\xrf@@n\hbox{{\rm(}$\@qn{#1}${\rm)}}}

% macro to make order use section and/or subsection numbers as a prefix

\def\n@@me#1#2>{#2}
\def\numberby#1{\xrf@@n
  \ifx\s@ction\undefined\else
  \expandafter\let\csname\s@@ve\endcsname=\s@ction\fi
  \ifx\subs@ction\undefined\else
  \expandafter\let\csname\subs@@ve\endcsname=\subs@ction\fi
  \numb@rby#1,>#1>}
\def\numb@rby#1,#2>#3>{\def\n@xt{#1}\ifx\n@xt\empty\taghe@d{}\else
  \def\n@xt{#2}\ifx\n@xt\empty\n@by#3>\else\n@@by#3>\fi\fi}
\def\n@by#1>{\ifx\s@cno\undefined\expandafter\expandafter
  \csname newcount\endcsname\csname s@cno\endcsname
  \csname s@cno\endcsname=0\else\s@cno=0\fi
  \xdef\s@@ve{\expandafter\n@@me\string#1>}
  \let\s@ction=#1\def#1{\global\advance\s@cno by 1
  \taghe@d{\number\s@cno.}\s@ction}}
\def\n@@by#1,#2>{\ifx\s@cno\undefined\expandafter\expandafter
  \csname newcount\endcsname\csname s@cno\endcsname
  \csname s@cno\endcsname=0\else\s@cno=0\fi
  \ifx\subs@cst\undefined\expandafter\expandafter
  \csname newcount\endcsname\csname subs@cst\endcsname
  \csname subs@cst\endcsname=0\else\subs@cst=0\fi
  \ifx\subs@cno\undefined\expandafter\expandafter
  \csname newcount\endcsname\csname subs@cno\endcsname
  \csname subs@cno\endcsname=0\else\subs@cno=0\fi
  \xdef\s@@ve{\expandafter\n@@me\string#1>}
  \let\s@ction=#1\def#1{\global\advance\s@cno by 1
  \global\subs@cno=\subs@cst
  \taghe@d{\number\s@cno.}\s@ction}
  \xdef\subs@@ve{\expandafter\n@@me\string#2>}
  \let\subs@ction=#2\def#2{\global\advance\subs@cno by 1
  \taghe@d{\number\s@cno.\number\subs@cno.}\subs@ction}}

% to reset counters

\def\numberfrom#1{\ifx\s@cno\undefined\else\n@mberfrom#1,>\fi}
\def\n@mberfrom#1,#2>{\def\n@xt{#2}%
  \ifx\n@xt\empty\n@@f#1>\else\n@@@f#1,#2>\fi}
\def\n@@f#1>{\s@cno=#1\advance\s@cno by -1}
\def\n@@@f#1,#2,>{\s@cno=#1\advance\s@cno by -1%
  \subs@cst=#2\advance\subs@cst by -1}

% macro to add additional prefix to numbers

\def\pret@g{}
\def\prefixby#1{\gdef\pret@g{#1}}

% now for the literature referencing macros - these differ because
% the reference generates the numbers (rather than the object being referenced)

% replace reference label by number

\newcount\r@fcount\r@fcount=0
\newcount\r@fcurr
\newcount\r@fmin
\newcount\r@fmax
\newcount\r@fone
\newcount\r@ftwo
\newif\ifc@te\c@tefalse
\newif\ifr@feat

\def\cite#1{{\rm[}\def\s@p{}\r@fmin=\r@fcount\r@fmax=0%
   \refn@te#1>>\r@fcurr=\r@fmin\advance\r@fcurr by-1\refc@te{\rm]}}

% first we make a note of which references are to be cited
% and give them a number if required
% the list is separated by commas, add one more and treat recursively

\def\refn@te#1>>{\refn@@te#1,>>}

\def\refn@@te#1,#2>>{\r@fnote{\str@pbl#1 >>}%
   \def\n@xt{#2}\ifx\n@xt\empty\else\refn@@te#2>>\fi}

\def\r@fnote#1%
  {\ifunc@lled{r@f}{#1}%
     \global\advance\r@fcount by 1\r@fmax=\r@fcount\r@fcurr=\r@fcount%
     \expandafter\xdef\csname r@f#1\endcsname{\number\r@fcount}%
     \expandafter\gdef\csname r@ftext\number\r@fcount\endcsname%
     {\message{ Reference #1 to be supplied }%
     Reference $#1$ to be supplied\par}%
   \else%
     \expandafter\r@fcurr=\csname r@f#1\endcsname\relax%
     \ifnum\r@fmin<\r@fcurr\else\r@fmin=\r@fcurr\advance\r@fmin by-1\fi%
     \ifnum\r@fmax<\r@fcurr\r@fmax=\r@fcurr\fi%
   \fi\expandafter\expandafter\def\csname r@fn\number\r@fcurr\endcsname{Y}}

\def\ifunc@lled#1#2{\expandafter\ifx\csname #1#2\endcsname\relax}

\def\str@pbl#1 #2>>{\str@@pbl#1#2 >>}
\def\str@@pbl#1 #2>>{\str@@@pbl#1#2 >>}
\def\str@@@pbl#1 #2>>{\str@@@@pbl#1#2 >>}
\def\str@@@@pbl#1 #2>>{#1}

% now loop through the range of possible cited numbers and cite,
% with appropriate punctuation

\def\refc@te{\r@featfalse\def\s@ve{}%
  {\loop\ifnum\r@fcurr<\r@fmax\advance\r@fcurr by 1\c@tefalse%
   \expandafter\refc@@te\number\r@fcurr>>%
   \ifc@te\expandafter\refe@t\number\r@fcurr>>\fi\repeat\s@ve}}

\def\refc@@te#1>>{\ifnum#1=0\edef\n@xt{}\else
   \edef\n@xt{\csname r@fn#1\endcsname}\fi%
   \expandafter\xdef\csname r@fn#1\endcsname{}%
   \ifx\n@xt\empty\else\relax\c@tetrue\fi}

\def\refe@t#1>>{\ifr@feat\ifnum\r@fone=\r@ftwo\res@cond#1>>%
   \else\reth@rd#1>>\fi\else\r@feattrue\ref@rst#1>>\fi}

\def\ref@rst#1>>{\r@feattrue\r@fone=#1\r@ftwo=#1%
   \s@p\expandafter\relax\number\r@fone}%

\def\res@cond#1>>{\advance\r@ftwo by 1\def\n@xt{#1}%
   \expandafter\ifnum\n@xt=\number\r@ftwo%
   \edef\s@ve{,\expandafter\relax\number\r@ftwo}%
   \else\def\s@p{,}\ref@rst#1>>\fi}%

\def\reth@rd#1>>{\advance\r@ftwo by 1\def\n@xt{#1}%
   \expandafter\ifnum\n@xt=\number\r@ftwo%
   \edef\s@ve{--\expandafter\relax\number\r@ftwo}\def\s@p{,}\else%
   \def\s@p{,}\s@ve\def\s@ve{}\ref@rst#1>>\fi}%

% read list of references and match with those cited

\def\refis#1 #2\par{\ifunc@lled{r@f}{#1}\else
   \expandafter\gdef\csname r@ftext\csname
r@f#1\endcsname\endcsname{#2\par}\fi}

% produce the references - check whether IOPP macros are defined already

\def\listreferences{%
\ifx\numrefjl\undefined
  \csname newdimen\endcsname\@ldskip\@ldskip=\parskip
  \csname newdimen\endcsname\@ldindent\@ldindent=\parindent
  \def\numr@f##1>>##2>>{\leftskip=40pt\parskip=0pt\parindent=-10pt%
    \hskip-30pt\hbox to 30pt{\rm[##1]\hss}##2\par\vskip\@ldskip%
    \parskip=\@ldskip\parindent=\@ldindent\leftskip=0pt}
  {\global\r@fcurr=0%
   \loop\ifnum\r@fcurr<\r@fcount\global\advance\r@fcurr by 1%
   \numr@f\number\r@fcurr>>\csname r@ftext\number\r@fcurr\endcsname>>\repeat}
\else
  {\global\r@fcurr=0%
   \loop\ifnum\r@fcurr<\r@fcount\global\advance\r@fcurr by 1%
   \def\refnum{\number\r@fcurr}\csname r@ftext\number\r@fcurr\endcsname\repeat}
\fi}

% the following only differs from plain TeX by not being outer

\def\beginsection#1\par{\vskip0pt plus.2\vsize\penalty-250
  \vskip0pt plus -.2\vsize\bigskip\vskip\parskip
  \leftline{\bf#1}\nobreak\smallskip\noindent}

\catcode`@=12
%
% Standard macros for setting out
% automatic punctuation is provided by the following.
%
\catcode`@=11
\newif\ifs@p
\def\jnlitem#1#2#3#4%
  {#1\def\l@st{#1}\ifx\l@st\empty\s@pfalse\else\s@ptrue\fi%
   \def\l@st{#2}\ifx\l@st\empty\else%
   \ifs@p, \fi{\frenchspacing\sl#2}\s@ptrue\fi%
   \def\l@st{#3}\ifx\l@st\empty\else\ifs@p, \fi{\bf#3}\s@ptrue\fi%
   \def\l@st{#4}\ifx\l@st\empty\else\ifs@p, \fi#4\s@ptrue\fi%
   \ifs@p.\fi\par}
\def\bkitem#1#2#3%
  {#1\def\l@st{#1}\ifx\l@st\empty\s@pfalse\else\s@ptrue\fi%
   \def\l@st{#2}\ifx\l@st\empty\else%
   \ifs@p, \fi{\frenchspacing\sl#2}\s@ptrue\fi%
   \def\l@st{#3}\ifx\l@st\empty\else\ifs@p, \fi#3\s@ptrue\fi%
   \ifs@p.\fi\par}
\catcode`@=12
%
% abbreviations for commonly quoted journals
%

\def\CMP{Comm. Math. Phys.}

\def\IJMPA{Int. J. Mod. Phys. A.}

\def\JPA{J. Phys. A: Math. Gen.}

\def\JSP{J. Stat. Phys.}

\def\NPB{Nucl. Phys. B}

\def\PD{Physica D}

\def\PLA{Phys. Lett. A.}

\def\PTP{Prog. Theor. Phys.}

%
% ---------------------------------------------------------------------
% Plain TeX, define page size and headings
% ---------------------------------------------------------------------
%
\magnification 1200
\hsize=15.6truecm
\vsize=23truecm
\newdimen\loffset\loffset=.6truecm
\newdimen\roffset\roffset=-.3truecm
\voffset=1truecm
\parskip=2pt
\nopagenumbers
\pageno=0
\headline={\ifnum\pageno=0{}\global\hoffset=\loffset\else \hdline\fi}
\def\hdline{\ifodd\pageno\rightheadline \else\leftheadline\fi}
\def\rightheadline{\tenrm\hfil{\it \rhead}\hfil\folio\global\hoffset=\loffset}
\def\leftheadline{\tenrm\folio\hfil{\it \lhead}\hfil\global\hoffset=\roffset}
%
% ---------------------------------------------------------------------
%
\catcode`@=11
\newif\ifnews@ct\news@ctfalse
\parskip=2pt plus 2pt
\font\twelvebf=cmbx10 scaled 1200
\font\twelveit=cmti10 scaled 1200
\newcount\secno\secno=0
\def\section#1#2\par#3\par
  {\vskip2cm\penalty-250\vskip-2cm\bigskip
  \if#1*\noindent{\bf#2}\else
  \advance\secno by 1\subsecno=0\news@cttrue
  \noindent\hbox to \parindent{\bf\number\secno.\hfil}{\bf#1}\fi
  \par\vskip-\parskip\medskip#3\news@ctfalse\par}
\newcount\subsecno\subsecno=0
\def\subsection#1
  {\advance\subsecno by 1
  \ifnews@ct\else\smallskip\fi
  \noindent\number\secno.\number\subsecno\hskip1ex{\sl #1.}\hskip1ex}
%
% ---------------------------------------------------------------------
% special characters which use gothic and black board bold fonts
% ---------------------------------------------------------------------
%
\ifx\Bbb\undefined
  \message{Black board bold font not found in macros.tex}
  \def\BC{{\bf C}}

\else
  \def\BC{{\Bbb C}}

\fi
\ifx\goth\undefined
  \message{Gothic font not found in macros.tex}
  \def\gs{{\sl s}}
  \def\gl{{\sl l}}
  
\else
  \def\gs{{\goth s}}
  \def\gl{{\goth l}}
  
\fi
\def\slt{\gs\gl_2}
\def\slth{\widehat{\slt}}
\def\U{U_q\bigl(\slt\bigr)}

\def\Uq{U_q\bigl(\slth\bigr)}
%
% ---------------------------------------------------------------------
% propositions, etc
% ---------------------------------------------------------------------
%
\order{defn}
\let\def@=\defn\let\refdef@=\refdefn
\def\defn#1#2.#3\enddef%
  {\csname proclaim\endcsname Definition \def@{#1}#2.#3\par\noindent}
\def\refdef#1{definition \refdef@{#1}}
\def\refDef#1{Definition \refdef@{#1}}
\order{lem}
\let\lem@=\lem\let\reflem@=\reflem
\def\lem#1#2.#3\endlem%
  {\csname proclaim\endcsname Lemma \lem@{#1}#2.#3\par\noindent}
\def\reflem#1{lemma \reflem@{#1}}
\def\refLem#1{Lemma \reflem@{#1}}
\order{prop}
\let\prop@=\prop\let\refprop@=\refprop
\def\prop#1#2.#3\endprop%
  {\csname proclaim\endcsname Proposition \prop@{#1}#2.#3\par\noindent}
\def\refprop#1{proposition \refprop@{#1}}
\def\refProp#1{Proposition \refprop@{#1}}
%
% ---------------------------------------------------------------------
% definifions for this paper
% ---------------------------------------------------------------------
%
\def\XXZ{{\rm X\kern-.1pt X\kern-.3pt Z}}
\def\CTM{{\rm C\kern-.3pt T\kern-.3pt M}}
\def\ABF{{\rm A\kern-.3pt B\kern-.4pt F}}
\def\VO{{V\kern-.5pt O}}
\def\geh{\goth{g}}
\def\ha{\goth{h}}

\def\Q{{\bf Q}}
\def\Z{{\bf Z}}
\def\B{B_n^{(1)}}
\def\D{D_n^{(1)}}

\def\A{A_n^{(1)}}
\def\F{{\cal F}}
\def\scR{{\cal R}}
\def\ep{\epsilon}
\def\vep{\varepsilon}
\def\Th{\Theta}
\def\th{\theta}
\def\la{\lambda}
\def\La{\Lambda}

\def\hc{\chk{h}}
\def\Rb{\overline{R}}

\def\Tc{\check T}

\def\br#1{\langle #1 \rangle}
\def\chk#1{#1^\vee}
\def\mod{~\hbox{mod}~}

\def\Hom{\mathop{\rm Hom}}
\def\End{\mathop{\rm End}}
\def\Im{\mathop{\rm Im}}
\def\Ker{\mathop{\rm Ker}}
\def\rank{\mathop{\rm rank}}
\def\id{\mathop{\rm id}}

\def\tr{\mathop{\rm tr}}

\def\Phit{\widetilde{\Phi}}
\def\Vd(#1){V^{(#1)*a^{\pm1}}}

\def\up{U'_q(\goth{g})}
\def\uq{U_q(\goth{g})}

\def\wt{\hbox{wt}\,}

\def\vac{|\hbox{vac}\rangle}

\def\ol#1{\overline#1}
\def\ot{\otimes}
\def\ket#1{|#1 \rangle}
%
%    Okado Macro
%
\def\W(#1,#2,#3,#4,#5,#6,#7,#8,#9)
{\hskip-4pt
        \pmatrix
        {
        #1&#2&#3\cr
        #4&&#5\cr
        #6&#7&#8\cr
        }
\hskip-4pt (#9)~}

\def\iso{{\buildrel \sim \over \rightarrow}}

\def\ch(#1,#2,#3){{#1\setminus\{#2\}\cup\{#3\}}}
%
%\Proof
%
\def\Proof{\noindent {\sl Proof.\quad}}
\def\Proofof#1{\noindent {\sl Proof of {#1}.\quad}}
%
%\Remark
%

%
%\Example
%

%
%\qed
%
\def\qed{\hbox{${\vcenter{\vbox{
    \hrule height 0.4pt\hbox{\vrule width 0.4pt height 6pt
    \kern5pt\vrule width 0.4pt}\hrule height 0.4pt}}}$}}
%
%Figure
%
\def\Figure(#1|#2|#3)
{\midinsert
\vskip #2
\hsize 9cm
\raggedright
\noindent
{\bf Figure #1\quad} #3
\endinsert}
%
%Table
%
\def\Table #1. \size #2 \caption #3
{\midinsert
\vskip #2
\hsize 7cm
\raggedright
\noindent
{\bf Table #1.} #3
\endinsert}
%
% ---------------------------------------------------------------------
% definifions for special use of \eqalignno and \leqalignno
% ---------------------------------------------------------------------
%
\expandafter\csname newskip\endcsname\c@ntering
\c@ntering=0pt plus 1000pt minus 1000pt
\def\eqalignsp#1{\displ@y \tabskip\c@ntering
  \halign to\displaywidth{\hfil$\@lign\displaystyle{##}$\tabskip\z@skip
    &$\@lign\displaystyle{{}##}$\hfil\tabskip\c@ntering
    &\llap{$\@lign##$}\tabskip\z@skip\crcr
    #1\crcr}}
\def\leqalignsp#1{\displ@y \tabskip\c@ntering
  \halign to\displaywidth{\hfil$\@lign\displaystyle{##}$\tabskip\z@skip
    &$\@lign\displaystyle{{}##}$\hfil\tabskip\c@ntering
    &\kern-\displaywidth\rlap{$\@lign##$}\tabskip\displaywidth\crcr
    #1\crcr}}
\def\okeqalignno#1{\xrf@@n\displ@y \tabskip=\c@ntering
  \halign to\displaywidth{\hfil$\displaystyle{##}$\tabskip=0pt
    &$\displaystyle{{}##}$\hfill\tabskip=1em
    &\hfil$\displaystyle{##}$\tabskip=0pt
    &$\displaystyle{{}\>##}$\hfill\tabskip=1em
    &$\displaystyle{{}##}$\hfill\tabskip=\c@ntering
    &\llap{$\eqaln@##$}\tabskip=0pt\crcr
    #1\crcr}}
%
% ---------------------------------------------------------------------
% cases with brace on RHS
% ---------------------------------------------------------------------
%
\def\rcases#1{\left.\,\vcenter{\normalbaselines\m@th
    \ialign{$##\hfil$&\quad##\hfil\crcr#1\crcr}}\right\}}
%
% ---------------------------------------------------------------------
%
\catcode`@=12
%
% ---------------------------------------------------------------------
%
\font\mb=cmmib10 at 12pt
\def\ltitle{Excitation Spectra of Spin Models constructed from \cr
            Quantized Affine Algebras of type
            $\hbox{\mb B}_n^{(1)}$, $\hbox{\mb D}_n^{(1)}$\cr}
\def\stitle{Excitation Spectra of Spin Models}

\def\lauthor{Brian Davies \cr
                 and\cr
             Masato Okado \cr}

\def\sauthor{Brian Davies  and Masato Okado}

\def\laddress{Department of Mathematics,
School of Mathematical Sciences, \hfil\break
Australian National University,
Canberra, ACT 0200, Australia. \hfil\break
{}\hfil\break
and\hfil\break
{}\hfil\break
Department of Mathematical Sciences,
Faculty of Engineering Science, \hfil\break
Osaka University,
Toyonaka, Osaka 560, Japan. \hfil\break}

\def\labstract{
The energy and momentum spectrum of the spin models constructed from the
vector representation of the quantized affine algebras of type $\B$ and
$\D$ are computed using the approach of Davies et al. \cite{DFJMN92}.
The results are for the anti-ferromagnetic (massive) regime, and they
agree with the mass spectrum found from the factorized S--matrix theory by
Ogievetsky et al. \cite{ORW87}.
The other new result is the explicit realization of the fusion
construction for the quantized affine algebras of type $\B$ and $\D$.}
%
% ---------------------------------------------------------------------
% Typeset first page and eject
% ---------------------------------------------------------------------
%
\gdef\lhead{\sauthor}
\gdef\rhead{\stitle}
\vglue 1cm
\centerline{\twelvebf\vbox{\halign{\hfil # \quad\hfil\cr\ltitle\crcr}}}
\vglue 1cm
\centerline{\twelveit\vbox{\halign{\hfil # \quad\hfil\cr\lauthor\crcr}}}
\vglue 2cm
\noindent\laddress
\vglue 1cm
\noindent June 1995
\vglue 1cm
\noindent 1991 {\it Mathematics subject classification}:
primary 82B13; secondary 82B20.
\vglue 1cm
{\narrower
\font\smr=cmr10 scaled 937
\font\smb=cmbx10 scaled 937
\noindent{\smb Abstract.}\hskip1ex\smr\labstract\par}
\vfill\eject
%
% ---------------------------------------------------------------------
% now for the actual body of the paper
% ---------------------------------------------------------------------
%
\numberby{\section}
%
%
%               Section 1
%
%
\section{Introduction}

In \cite{DFJMN92} was given a new scheme for solving the
six-vertex model and associated $\XXZ$ chain, in the antiferromagnetic
regime, using the newly discovered quantum affine symmetry of the
system.
The approach of that paper has been extended to
higher spin chains \cite{IIJMNT92}, to the higher rank case \cite{DO93}
and to the \ABF\ models of Andrews, Baxter and Forrester \cite{JMO92}.
All of these papers are concerned with models constructed on the
(quantised affine) algebra $A^{(1)}_n$.
In this paper we treat the case that the algebra is either $\B$ or
$\D$.
(The bosonization of level 1 vertex operators for $\D$ and an
integral formula for the correlation functions of the vertex model are
given in \cite{Ko}.)
Our principal result is to identify the ``particle spectrum'',
including their excitation energies and dispersion relations.
We shall see also that the mass spectrum of these particles coincides with
the mass spectrum found by Ogievetsky, Reshetikhin and Wiegmann from
the factorised S--matrix theory \cite{ORW87}.

The ideas are fundamentally different from the long established method of
the Bethe Ansatz.
One of the questions on Feynman's last blackboards
reads ``describe centre of string without the ends'' \cite{F89}.
In the Bethe Ansatz approach, the answer is to join the ends
and then take the infinite limit.
With the scheme first presented in \cite{DFJMN92}, the system is
always infinite: one selects an arbitrary point as the centre and treats
the system as the union of its left and right hand parts.
The role played by the ends is then simply to select the various ground
states.

To assist in the presentation of this paper, which has many technical
details, we first make some general introductory comments in the present
section, using the example of the \XXZ\ chain.
Then, in section 2, we discuss more specifically the application of the
method to the problem at hand and state the principal results.

\subsection{Role of corner transfer matrices}
Corner transfer matrices (\CTM s) were invented by Baxter
\cite{Bax76,Bax77}, and proved to be an effective method for the
evaluation of one-point functions in exactly solved lattice models
of statistical mechanics \cite{Bax80,ABF84,JMO88b,DJKMO87}.
\CTM s are defined as a suitably normalised partition function of a
whole quadrant of the lattice, so they are a transfer matrix which
acts on a semi-infinite spin chain.
{}From an early stage it was known that they have the following remarkably
simple properties:
\par\noindent(i) the spectrum is integer powers of a single
parameter;
\par\noindent(ii) they are exactly an exponential
$A(u)=\exp(uK)$, where $K$ is a spin-chain operator of the general form
$K=\sum_{l=1}^\infty lH_l$, and $u$ is the (additive) spectral parameter;
\par\noindent(iii) the relation to more usual Hamiltonian $H$ associated
with the row transfer matrix $T(u)$ is $H=\sum_{l=-\infty}^\infty H_l$;
\par\noindent(iv) this relation is manifest as a ``boost property''
$T(u)A(v)=A(v)T(u+v)$ \cite{SW83,Th86}.

Since the transfer matrix $T(u)$ takes a simple form (typically a shift
operator) when $u=0$, one sees that a full description of the row
transfer matrix, and the Hamiltonian $H$, may be had if one can understand
the eigenstates of \CTM s and can also construct the shift operator in
that representation.
The vital key is the connection with quantum affine algebras
\cite{FM92,Dav93,Dav94}.
In this picture, the six-vertex \CTM\ generator $K$, acting on the
left-hand semi-infinite part, is identified with the derivation operator
$d$ of the quantised affine algebra $\Uq$, and its eigenstates with the
weight vectors of the level $1$ modules $V(\Lambda_0)$, $V(\Lambda_1)$.
The two highest weights $\Lambda_i$ correspond to the
two possible ground states for the \XXZ\ model.
The eigenstates of the right hand part is the dual (a level $-1$ module);
so a suitable representation for the space of states of the entire chain
is the direct sum of tensor products $\bigoplus_{i,j=0,1}
V(\Lambda_i)\otimes V(\Lambda_j)^{*a}$  --- a
level $0$ module.
(The antipode $a$ must be used to construct the dual; the definitions are
given below.)
Unlike the irreducible level $1$ modules, this
representation is highly reducible: the most obvious reduction is a
decomposition into $n$-particle states.

\subsection{Translation operator}
One may shift the selected point at which translational symmetry is
broken, one site at a time, using the theory of quantum vertex operators
(\VO s) due to Frenkel and Reshetikhin \cite{FR92}.
This gives a viable representation-theoretic realisation of
the translation operator $T=T(0)$.
Since the derivation $d$ was already identified with a Hamiltonian
spin chain (the CTM) whose coefficients are linear in the
position along the chain, the usual Hamiltonian spin chain becomes
identified with a multiple of the operator $(TdT^{-1}-d)$, using the
boost property.

\VO s are algebra homomorphisms between highest weight
modules $V(\Lambda_i)$ ($i=0,1$) and tensor products of the form
$V(\Lambda_{1-i})\otimes V_z$ or $V_z\otimes V(\Lambda_{1-i})$.
($V_z$ is a loop module, or affinisation of the usual spin-half $\U$
module.)
For the two different orders of tensor product, we call the corresponding
\VO s type I and type II, respectively.
A type I \VO\ expands the eigenstates of the left-hand semi-infinite part
in terms of the state of one local site and the eigenstates of a new
semi-infinite part, truncated by that site.
This ``splits off'' one site at the centre of the string by mapping
$V(\Lambda_i)$ to $V(\Lambda_{1-i})\otimes V_z$.
By using duality properties of modules and homomorphisms between them,
there is another type I \VO\ which adds this
single site to the right hand semi-infinite part, that is, which maps
$V_z\otimes V(\Lambda_i)^{*a}$ to $V(\Lambda_{1-i})^{*a}$.
The middle of the string is thereby translated by one lattice unit,
whilst the  sectors are switched corresponding to the translation.
The advantage of this construction is that all expansions are convergent
in $q$ and $z$, in some domain of analyticity, even though they have
rather complicated forms.
Moreover, closed form expressions may be obtained for physical quantities
using the powerful machinery of modern algebra and representation theory.

\subsection{Creation and annihilation operators}
Consider the ground states.
Since the \XXZ\ chain is $\Uq$ invariant in this infinite limit, the
ground states $\Psi_i$ ($i=0,1$) must be unique, that is, $\Psi_i$ must span a
one-dimensional (and therefore trivial) sub-representation.
Using the canonical identifications
$$
V(\Lambda_i)\otimes V(\Lambda_i)^{*a}\quad
\mathop{\longrightarrow}^\sim\quad
\hbox{$\Hom$}_\BC(V(\Lambda_i),V(\Lambda_i)),
$$
we see that the ground states are simply identity maps of $V(\Lambda_i)$
into itself.
This gives a simple expansion for $\Psi_i$ in terms of any dual basis
system for $V(\Lambda_i)$ and $V(\Lambda_i)^{*a}$.

To create particles from a ground state is equivalent to finding
maps $\varphi^*_\pm(z)$ which create sub-modules in the various sectors
$V(\Lambda_i)\otimes V(\Lambda_j)^{*a}$.
For a single particle of spin $1/2$ and momentum
$z={\rm e}^{ip}$, the submodule must be isomorphic to the spin-half $\Uq$
module $V_z$, so we seek a homomorphism
$$
\varphi^*_\pm(z):V_z\otimes V(\Lambda_i)\quad
\mathop{\longrightarrow}^{U_q}\quad V(\Lambda_{1-i})
$$
for the creation operator.
Using the canonical identification
$$
\hbox{$\Hom$}_{U_q}(L\otimes M,N)
\quad\mathop{\longrightarrow}\limits^\sim\quad
\hbox{$\Hom$}_{U_q}(M,L^{*a^{-1}}\otimes N)
$$
one finds that the creation operators are equivalent to \VO s of type II:
$$
V(\Lambda_i)
\quad\mathop{\longrightarrow}^{U_q}\quad
V_z^{*a^{-1}}\otimes V(\Lambda_{1-i}).
$$

For the annihilation operators, we seek the following \VO s of type II:
$$
V(\Lambda_i)
\quad\mathop{\longrightarrow}^{U_q}\quad
V_z\otimes V(\Lambda_{1-i}).
$$

\subsection{Selection rules}
For the \XXZ\ model, and also the higher spin \cite{IIJMNT92} and higher rank
\cite{DO93} cases, the excitation energies, particle spectra and dispersion
relations are compared with earlier Bethe Ansatz calculations.
These are important checks on the validity of the representation-theoretic
picture versus the physical content.
But it is equally important to note that the information, including all
necessary commutation relations, scalar factors and selection rules, may
be obtained directly from representation theory.
As a simple example, consider again the \XXZ\ model.
The question arises, why is there only a spin-half particle in the spectrum?
The answer from representation theory is that, if one considers the tensor
product $V^s_z\otimes V(\Lambda_i)$, where $s$ is spin, there is no
\VO\ which maps this into the space $V(\Lambda_j)$
except when $s=1/2$ \cite{DJO93}.
The language of quantum \VO s may be rather new to mathematical physics,
but selection rules are certainly not.
Similarly, for the higher rank case $A^{(1)}_n$, all the level $1$ modules
$V(\Lambda_i)$, $i=0,\ldots,n$, correspond to spaces of states with
differing boundary conditions.
For this case, let $V_z^{(k)}$, $k=2,\ldots,n$, be the $k$-th fusion of the
(affinised) vector representation $V_z^{(1)}$; then it is known that there is
a  \VO\ which maps each $V_z^{(k)}\otimes V(\Lambda_i)$ into a
unique $V(\Lambda_j)$, so that spectrum has a total of $n$ different
particle types.
%
%
%               Section 2
%
%
\section{Main Results}

\subsection{Hamiltonian and the anti-ferromagnetic regime}
We consider affine Lie algebras
$\geh=B_n^{(1)}$, $(n\ge3)$ (resp. $D_n^{(1)}$, $(n\ge4)$).
Their Dynkin diagrams are shown below.
\midinsert

\vbox{\vskip 0.8cm
\centerline{\epsfxsize=11cm\epsfbox{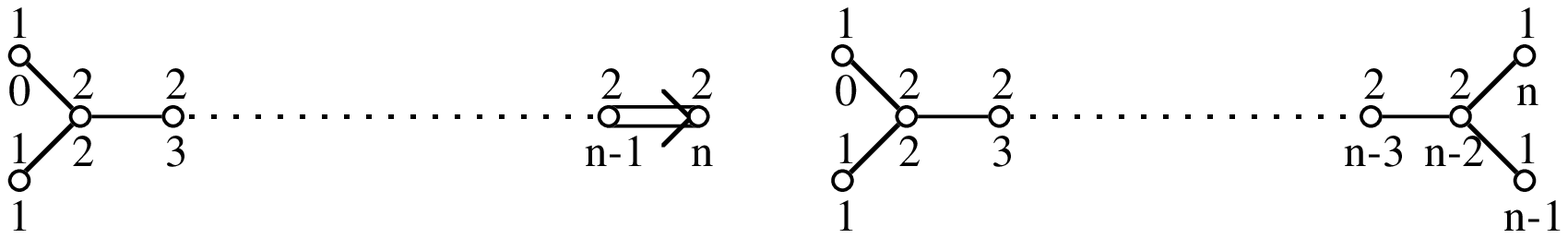}}
\vskip 0.5cm
\centerline{Figure 1: Dynkin diagrams of $\B$ and $\D$}
\vskip 0.5cm}
\endinsert

Let $\uq$ be the quantum affine algebra associated with $\geh$.
Consider the $R$--matrix $R(z)=\Rb^{(1,1)}(z)$ \ref{Rvec} for the
vector representation of $\uq$. $R(z)$ acts on the vector space
$V\ot V$, where $\dim V=2n+1$ (resp. $2n$).
Given the $R$-matrix, it provides the Boltzmann weights
of a vertex model with $(2n+1)$  (resp. $2n$) states, satisfying the
Yang-Baxter equation \ref{YBE}.
Set
$$
\eqalignno{
h&=-(q^t-q^{-t})\cdot z{d\over dz}\log PR(z)\biggr|_{z=1},\cr
t&=1/2~(\geh=\B),\quad t=1~(\geh=\D).
&(deft)\cr}
$$
Here $P$ denotes the transposition: $Pu\ot v=v\ot u$.
Define the Hamiltonian
$$
{\cal H}=\sum_{l\in\Z}h_{l+1,l}
$$
acting on the infinite tensor product
$$
\matrix{
\cdots&\ot& V&\ot& V&\ot& V&\ot&\cdots.\cr
&&{\scriptstyle l+1}&&{\scriptstyle l}&&{\scriptstyle l-1}&&\cr}
$$
Here $h_{l+1,l}$ acts as $h$ on the $(l+1)$-th and $l$-th components
and as the identity on the other components.

The structure of space of states (physical regime) depends on the region of the
parameter $q$ in the Hamiltonian.
Unlike the Bethe Ansatz method, the present approach is effective only for a
particular region of $q$, corresponding to the anti-ferromagnetic or massive
regime.
It is in this regime that the necessary infinite limit of the physical
system may be defined in such a way that the \CTM\ becomes identified with the
derivation of the quantum affine algebra.
Because of the large freedom of gauge transformations of the Hamiltonian, we do
not attempt herein to investigate this problem along the lines of
\cite{FM92,Dav93}, although we hope to do so in a future publication.
(In fact, even the general $A_n^{(1)}$ case has not been analysed in
such a way, although the agreement with existing Bethe Ansatz results provides
a
strong check in that case.)
Here we just state our conjecture, which is a simple extension of known
results, and takes account of the isomorphisms between the $A^{(1)}_1$ and
$B^{(1)}_1$ case and the $A^{(1)}_3$ and $D^{(1)}_3$ case.
It is
$$
-1<q^t<0,
$$
where $t$ is given in \ref{deft}.
The prefactor $-(q^t-q^{-t})$ of $h$ is chosen to conform to conventions set in
previous work which relates quantum affine algebras and \CTM s.
With this choice, it is negative.

\subsection{Physical sectors and the space of states}
Physical sectors are the eigenstates of the \CTM\ with some
appropriate boundary condition. Candidates for them are integrable
highest weight modules.
The situation here is quite different from the $A^{(1)}_n$ case.
There, all the highest weight representations corresponding to the
fundamental weights $V(\Lambda_i)$,
$i=0,\ldots,n$ are of level $1$.
Since the affinised modules used to construct translation and creation
operators are of level $0$, all of the fundamental modules are involved in
the physics, whilst all of the higher level modules are excluded.
For the orthogonal algebras, the fundamental modules $V(\Lambda_0)$,
$V(\Lambda_1)$, $V(\Lambda_n)$ (also $V(\Lambda_{n-1})$ for $\D$) are
level $1$ whilst all others are level $2$.
For the anti-ferromagnetic regime to which we restrict our attention, there
are therefore three (resp. four) \CTM\ ground state sectors, corresponding
to these level-$1$ modules.
Therefore, the whole physical space of states may be identified with
$$
\bigoplus_{i,j}V(\La_i)\ot V(\La_j)^{*a},
$$
where the sum is over $0,1,n$ (also $n-1$ for $\D$).
As a $\uq$-module, the sector  $V(\Lambda_i)\otimes V(\Lambda_j)^{*a}$ is
canonically equivalent to the space of linear maps
$V(\Lambda_j)\mathop{\longrightarrow}\limits^\BC V(\Lambda_i)$,
and the ground states of the infinite chain are the identity maps
which require $i=j$.

\subsection{Selection rules}
Let $V_z^{(1)}$ be the affinised vector representation of $\B$
(resp. $\D$), and let $V_z^{(k)}$ be the affinised $k$'th fusion, with
$k=2,\ldots,n-1$  (resp. $k=2,\ldots,n-2$).
We shall also need to consider the affinised spin representations, which we
denote here as $V_z^{(n)}$ (also $V_z^{(n-1)}$ for $\D$).
To each of the representations $V_z^{(k)}$ there correspond type II \VO s
between physical sectors, and therefore massive excitations of the
system for each $1\le k\le n$.
We shall describe the selection rules in detail.
This will provide a proper setting for the excitation spectra formulae which
follow, and give a comprehensive picture of how the various sectors are
inter-related.
First, for the vector representation $V_z^{(1)}$ we have the following
\VO s (the two columns are for $\B$ and $\D$ as indicated):
$$
\eqalign{
&\B \cr
V_z^{(1)}\otimes V(\Lambda_0)&~
\longrightarrow~V(\Lambda_1), \cr
V_z^{(1)}\otimes V(\Lambda_1)&~
\longrightarrow~V(\Lambda_0), \cr
V_z^{(1)}\otimes V(\Lambda_n)&~
\longrightarrow~V(\Lambda_n). \cr
\phantom{V_z^{(1)}\otimes V(\Lambda_n)} \cr}\qquad
\eqalign{
&\D \cr
V_z^{(1)}\otimes V(\Lambda_0)&~
\longrightarrow~V(\Lambda_1), \cr
V_z^{(1)}\otimes V(\Lambda_1)&~
\longrightarrow~V(\Lambda_0), \cr
V_z^{(1)}\otimes V(\Lambda_n)&~
\longrightarrow~V(\Lambda_{n-1}), \cr
V_z^{(1)}\otimes V(\Lambda_{n-1})&~
\longrightarrow~V(\Lambda_n). \cr}
\eqno(sel1)
$$
So, for example, the creation operator corresponding to $V_z^{(1)}$, acting on
the ground state in the sector $V(\Lambda_0)\otimes V(\Lambda_0)^{*a}$,
creates a massive excitation in the sector $V(\Lambda_1)\otimes
V(\Lambda_0)^{*a}$, and acting on the ground state in sector
$V(\Lambda_n)\otimes V(\Lambda_n)^{*a}$ it creates an excitation
in the sector $V(\Lambda_n)\otimes V(\Lambda_n)^{*a}$ for $\B$ or
$V(\Lambda_{n-1})\otimes V(\Lambda_n)^{*a}$ for $\D$.

For the fusion representations, the selection rules are exactly the
same as one would find by iterating the vector creation operator.
This is consistent with the fact that $V^{(k)}$ is projected out of
$(V^{(1)})^{\otimes k}$.
Thus, when $k$ is odd, the picture is the same as \ref{sel1}, whereas when
$k$ is even we have
$$
V_z^{(k)}\otimes V(\Lambda_i)
\longrightarrow~V(\Lambda_i)
\qquad\cases{
\hbox{$(i=0,1,n)$\phantom{$,n-1$} for $\B$},\cr
\hbox{$(i=0,1,n-1,n)$ for $\D$}.\cr}
\eqno(sel2)
$$

The remaining excitations come from the spin representations $V_z^{(n)}$
(and  $V_z^{(n-1)}$ for $\D$): these are the only ones which can mix
$V(\Lambda_0)$ and $V(\Lambda_1)$ with $V(\Lambda_n)$ (and  $V(\Lambda_{n-1})$
for $\D$).
The possibilities are:
$$
\eqalign{
&\qquad\qquad\B \cr
V_z^{(n)}&\otimes V(\Lambda_j)~
\longrightarrow~V(\Lambda_n), \cr
V_z^{(n)}&\otimes V(\Lambda_n)~
\longrightarrow~V(\Lambda_j), \cr
&\phantom{V_z^{(n)}} \cr
&\phantom{V_z^{(n)}} \cr}\qquad
\eqalign{
&\qquad\qquad\D \cr
V_z^{(n)}&\otimes V(\Lambda_j)~
\longrightarrow~V(\Lambda_{n-j}), \cr
V_z^{(n-1)}&\otimes V(\Lambda_j)~
\longrightarrow~V(\Lambda_{n+j-1}), \cr
V_z^{(n-i)}&\otimes V(\Lambda_{n-j})~
\longrightarrow~V(\Lambda_j), \cr
V_z^{(n+i-1)}&\otimes V(\Lambda_{n-j})~
\longrightarrow~V(\Lambda_{1-j}), \cr}
\eqno(sel3)
$$
where $j$ takes the values $0,1$, and in the $\D$ case $i=0$ (resp. $i=1$) if
$n$ is even (resp. $n$ is odd).

\subsection{Creation/annihilation operators and the transfer matrix}
As we have stated in the introduction, type II \VO s are used for the
mathematical formulations of creation and annihilation operators. Let
us denote them by $\varphi_{\la,I}^{(k)*}(z)$ and $\varphi_{\la,I}^{(k)}(z)$.
Here $k$ is related to the affinised representation $V_z^{(k)}$ we specify
for the type II \VO. $\la$ is the ground state sector they act, $I$ is
an index of base vectors of $V_z^{(k)}$, and $z$ is the momentum. See
subsection 7.3 for the precise definition. For a level 1 dominant integral
weight
$\la$, {\it i.e.} $\la=\La_i$ with $i=0,1,n$ (and $n-1$ for $\D$), define
$\la^{(k)*}$, $\la^{(k)}$ ($k=1,\cdots,n$) to be the one determined
uniquely from the following selection rules:
$$
\eqalign{
&V_z^{(k)}\ot V(\la)\longrightarrow V(\la^{(k)*}),\cr
&V(\la)\longrightarrow V_z^{(k)}\ot V(\la^{(k)}).\cr}
$$
We can see $\left(\la^{(k)}\right)^{(k)*}=\left(\la^{(k)*}\right)^{(k)}
=\la$. $\varphi_{\la,I}^{(k)*}(z)$ and $\varphi_{\la,I}^{(k)}(z)$ are
operators from $V(\la)$ to $V(\la^{(k)*})$ and $V(\la^{(k)})$, respectively.

The row transfer matrix can also be formulated via {\VO} of type I
(See subsection 7.3):
$$
T_{\la\mu}^{\la^{(1)}\mu^{(1)}}(z):\quad
V(\la)\ot V(\mu)^{*a}\longrightarrow V(\la^{(1)})\ot V(\mu^{(1)})^{*a}.
$$
{}From the commutation relations of \VO s, we can derive those between
the transfer matrix and creation (annihilation) operators. They are
given by
$$
\eqalign{
\varphi_{\la^{(1)},I}^{(k)*}(z_1)T_{\la\mu}^{\la^{(1)}\mu^{(1)}}(z_2)
&=\tau^{(k)}(z_1/z_2)
T_{\la^{(k)*}\mu}^{(\la^{(k)*})^{(1)}\mu^{(1)}}(z_2)\
\varphi_{\la,I}^{(k)*}(z_1),\cr
\varphi_{\la^{(1)},I}^{(k)}(z_1)T_{\la\mu}^{\la^{(1)}\mu^{(1)}}(z_2)
&=\tau^{(k)}(z_1/z_2)^{-1}
T_{\la^{(k)}\mu}^{(\la^{(k)})^{(1)}\mu^{(1)}}(z_2)\
\varphi_{\la,I}^{(k)}(z_1).\cr}
$$
Here $\tau^{(k)}(z)$ is a scalar function defined in the next subsection.

\subsection{Excitation spectra}
Because of the number of possible ground state sectors, there is considerable
detail in the selection rules for massive particle excitations.
By contrast, the excitation spectra are quite simple.
Regardless of the sectors, there is just a single dispersion relation for each
value of $1\le k\le n$: moreover there is no distinction between
$k=n-1$ and $k=n$ in the $\D$ case.
Here we discuss briefly the formulae for the $\A$, $\B$ and $\D$ cases, and
compare with the mass spectrum found by Ogievetsky, Reshetikhin and  Wiegmann
from the factorised S--matrix theory \cite{ORW87}. We note here that
Nakanishi \cite{Na} gave a systematic way to obtain the mass spectrum via
the Yangian algebra symmetry.

Let us first discuss the $\A$ case, which was treated by Date and Okado
\cite{DO93} using the quantum affine symmetry.
It was shown there that the excitation spectrum is given as
$$
{\rm e}^{-ip^{(k)}(\theta)}=\tau^{(k)}(z),\qquad
\epsilon^{(k)}(\theta)=-(q-q^{-1})z{d\over dz}\log\tau^{(k)}(z),
\eqno(ept)
$$
where $\epsilon^{(k)}(\theta)$, $p^{(k)}(\theta)$ are the energy and momentum
with rapidity variable $\theta$ ($-z=e^{2i\theta}$), and
$$
\eqalignno{
\tau^{(k)}(z)&=z^{{k\over n+1}-1}
 {\Theta_{q^{2n+2}}(-(-q)^kz)\over\Theta_{q^{2n+2}}(-(-q)^kz^{-1})},&(aex)\cr
\Theta_p(z)&=(z;p)_\infty(pz^{-1};p)_\infty(p;p)_\infty,&(theta)\cr
(z;p)_\infty&=\prod_{j=0}^\infty(1-p^jz).&(prod)\cr}
$$
Note that the affine Lie algebra of type $A_n^{(1)}$ is
$\widehat{\goth{sl}}_{n+1}$, so that $n$ in \cite{DO93} should be
shifted by 1.

Date and Okado also show that this is exactly the result previously obtained by
Babelon, deVega and Viallet \cite{BDV83} from the Bethe Ansatz.
The latter also consider the scaling limit of small lattice spacing and small
rapidity, to obtain the relativistic spectrum
$$
P^{(k)}(\theta)=\mu\sin\left({\pi k\over n+1}\right){\rm sh}\,v,\qquad
E^{(k)}(\theta)=\mu\sin\left({\pi k\over n+1}\right){\rm ch}\,v.
$$
Here $P^{(k)}$, $E^{(k)}$ and $v$ are appropriately scaled version of
$p^{(k)}$, $\epsilon^{(k)}$ and $\theta$.
The mass spectrum can be read off directly from this:
$$
m^{(k)}=\mu\sin\left({\pi k\over n+1}\right),\qquad
k=1,\ldots,n.
$$
This agrees with the result obtained in \cite{ORW87,Na}.

For the present calculations, we find herein that the excitation spectrum
is given via the
following functions, in the cases that the excitation arises from the fusion
or spin representation:
$$
\eqalignsp{
\tau^{(k)}(z)&=z^{-1}{\Th_{\xi^2}(-(-q)^k z)\Th_{\xi^2}(-(-q)^{-k}\xi z)
              \over\Th_{\xi^2}(-(-q)^k z^{-1})
                   \Th_{\xi^2}(-(-q)^{-k}\xi z^{-1})},
              &\hbox{(fusion)}\hskip.5cm\cr
              &=z^{-1/2}{\Th_{\xi^2}(-s\xi^{1/2}z)
              \over\Th_{\xi^2}(-s\xi^{1/2}z^{-1})}.
              &\hbox{(spin)}\hskip.5cm\cr}
$$
Here $\xi=q^{\hc}$ where $\hc$ is the dual Coxeter number of $\geh$
($2n-1$ for $\B$, $2n-2$ for $\D$), and $s=(-)^n$ ($\geh=\B$), $s=(-)^{n-1}$
($\geh=\D$).
For the fusion excitations, $k$ takes the values $1\le k\le n-1$ ($\geh=\B$)
or $1\le k\le n-2$ ($\geh=\D$) ($k=1$ is the vector representation); the
remaining values of $k$ are for the spin excitations. (The authors could not
find any published Bethe Ansatz result to compare with. But J. Suzuki sent
M.O. a fax showing that, apart from overall sign, our results agree with
those computed from their Bethe Ansatz data in \cite{KNS94}.)
If we apply a similar scaling limit as in the $\A$ case, we obtain
$$
\eqalign{
&\hbox{fusion}\cr
P^{(k)}(\theta)&=2\mu\sin\left({\pi k\over \hc}\right){\rm sh}\,v,\cr
E^{(k)}(\theta)&=2\mu\sin\left({\pi k\over \hc}\right){\rm ch}\,v.\cr}
\hskip-.7cm\eqalign{
&\hbox{spin}\cr
\phantom{\left({\pi k\over N-2}\right)}
P^{(k)}(\theta)&=\mu\,{\rm sh}\,v,\cr
\phantom{\left({\pi k\over N-2}\right)}
E^{(k)}(\theta)&=\mu\,{\rm ch}\,v.\cr}
$$
The mass spectrum agrees with that obtained in \cite{ORW87,Na}.
An interesting feature is the factor $2$ which occurs for the fusion
excitations; this comes from the double $\Theta_{\xi^2}$ factors in
$\tau^{(k)}(z)$.

The excitation spectrum is the main physical result of this paper.
However, the remaining sections, and the appendix, which are devoted to the
derivation of these results, contain a considerable amount of new information
about the $R$-matrices for the quantised affine algebras of type $\B$ and $\D$.
In particular, the fusion construction is treated in some detail.
%
%
%               Section 3
%
%
\section{Fundamental representations and $R$--matrices}

\subsection{Quantum affine algebras}
We consider affine Lie algebras $\geh=\B$ $(n\ge3)$,
$\D$ $(n\ge4)$.
Let $\ha^*,\La_i,h_i=\chk{\alpha}_i,\alpha_i,\delta=\sum_{i=0}^{n}a_i\alpha_i$
and $d$ have the same meaning as in \cite{Kac90}.
Their Dynkin diagrams are shown in Figure 1.
The lower (resp. upper) integer of each vertex stands for the index $i$
(resp. the value $a_i$).
The dual Coxeter number is defined as $\hc=\sum_{i=0}^n
\chk{a}_i$, where $\chk{a}_i=a_i$ except that for $\geh=\B \chk{a}_n=a_n/2$,
so that we have $\hc=2n-1~(\geh=\B)$, $2n-2~(\geh=\D)$.
The invariant bilinear form $(\,|\,)$ in \cite{Kac90} is so normalised that
$(\theta|\theta)=2$,
where $\theta=\delta-\alpha_0$. We put $\rho=\sum_{i=0}^n\La_i$.
For $\la\in{\ha}^*$ $\ol{\la}$ denotes the orthogonal projection of $\la$ on
$\ol{{\ha}}^*$, where $\ol{{\ha}}^*$ is the linear span of the classical roots
$\alpha_1,\cdots,\alpha_n$. Following \cite{Bou} we introduce an orthonormal
basis $\{\ep_1,\cdots,\ep_n\}$ of $\ol{{\ha}}^*$, by which $\alpha_i,
\ol{\La}_i,\ol{\rho}$ are represented below.
$$
\eqalignno{
\alpha_i&=\ep_i-\ep_{i+1}\hskip2.1cm(1\le i\le n-1),\cr
&=\ep_n\hskip3.06cm(i=n\hbox{ for }\B),\cr
&=\ep_{n-1}+\ep_n\hskip1.95cm(i=n\hbox{ for }\D),\cr
\ol{\La}_i&=\ep_1+\cdots+\ep_i
\hskip1.6cm(1\le i\le n-1\hbox{ for }\B,1\le i\le n-2\hbox{ for }\D),\cr
&={\ep_1+\cdots+\ep_{n-1}-\ep_n \over2}\quad(i=n-1\hbox{ for }\D),\cr
&={\ep_1+\cdots+\ep_{n-1}+\ep_n \over2}\quad(i=n),\cr
2\ol{\rho}&=(2n-1)\ep_1+(2n-3)\ep_2+\cdots+3\ep_{n-1}+\ep_n
\quad(\hbox{ for }\B),\cr
&=(2n-2)\ep_1+(2n-4)\ep_2+\cdots+2\ep_{n-1}\quad(\hbox{ for }\D).\cr}
$$

For our affine Lie algebras there are Dynkin diagram automorphisms described
in Figure 2.
We shall use these symmetries to reduce the number of cases of two point
functions which must be calculated.
%***

\midinsert
\vbox{\vskip 0.8cm
\centerline{\epsfxsize=11cm\epsfbox{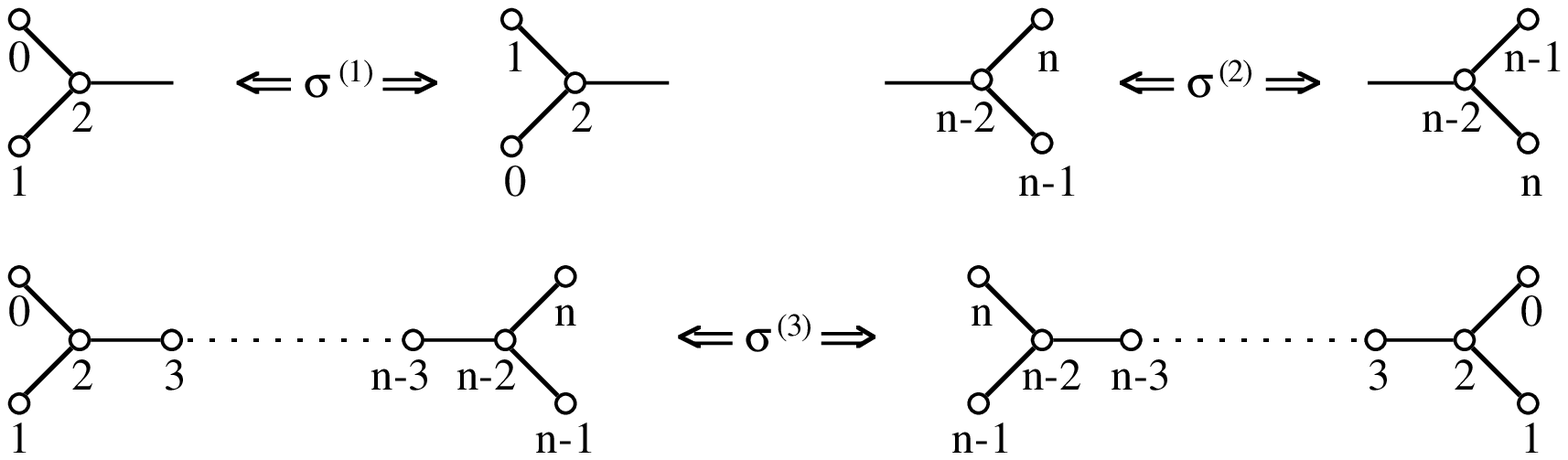}}
\vskip 0.5cm
\centerline{{\bf Figure 2:} Dynkin diagram automorphisms}
\vskip 0.5cm}
\endinsert

The quantum affine algebra $U_q(\geh)$ is defined as the $\Q(q)$--algebra
with 1 generated by $e_i,f_i,t_i(i=0,\cdots,n),q^d$ satisfying
$$
\eqalignno{
&[t,t']=0\hskip2.1cm\hbox{for}\quad t,t'\in\{t_0,\cdots,t_n,q^d\},\cr
&t_ie_jt_i^{-1}=q^{(\alpha_i|\alpha_j)}e_j,\quad
t_if_jt_i^{-1}=q^{-(\alpha_i|\alpha_j)}f_j,\cr
&q^de_jq^{-d}=q^{\delta_{j0}}e_j,\,\qquad
q^df_jq^{-d}=q^{-\delta_{j0}}f_j,\cr
&[e_i,f_j]=\delta_{ij}(t_i-t_i^{-1})/(q_i-q_i^{-1}),\cr
&\sum_{k=0}^b(-)^ke_i^{(k)}e_je_i^{(b-k)}
=\sum_{k=0}^b(-)^kf_i^{(k)}f_jf_i^{(b-k)}=0\quad(i\ne j),\cr}
$$
where $b=1-\br{h_i,\alpha_j}$. Here we have set $q_i=q^{(\alpha_i|\alpha_i)/2},
[m]_i=(q_i^m-q_i^{-m})/(q_i-q_i^{-1}),[k]_i!=\prod_{m=1}^k[m]_i,e_i^{(k)}=
e_i^k/[k]_i!,f_i^{(k)}=f_i^k/[k]_i!$.
We denote by $U'_q(\geh)$ the subalgebra of
$U_q(\geh)$ generated by the elements $e_i,f_i,t_i$ ($i=0,\cdots,n$).
Let $x_i$ be any of $e_i,f_i,t_i$. We define algebra automorphisms
$\sigma^{(1)},
\sigma^{(2)},\sigma^{(3)}$ of $U'_q(\geh)$ as follows (Note that
$\sigma^{(2)},\sigma^{(3)}$ are only for $\D$.):
$$
\eqalign{
&\sigma^{(1)}(x_i)=x_{1-i}~(i=0,1),\cr
&\sigma^{(2)}(x_i)=x_i~(i\le n-2),\cr
&\sigma^{(3)}(x_i)=x_{n-i}(\forall i).\cr}
\qquad
\eqalign{
&\sigma^{(1)}(x_i)=x_i~(i\ge2),\cr
&\sigma^{(2)}(x_{n-i})=x_{n+i-1}~(i=0,1),\cr
&\phantom{\sigma^{(3)}(x_i)}\cr}
\eqno(Dda)
$$

For a representation $(\pi,V)$ of $U'_q(\geh)$, we put $V_z=V\ot\Q(q)
[z,z^{-1}]$, and lift $\pi$ to a representation $(\pi_z,V_z)$ of $U_q(\geh)$
as follows:
$$
\eqalignno{
\pi_z(e_i)(v\ot z^n)&=\pi(e_i)v\ot z^{n+\delta_{i0}},\cr
\pi_z(f_i)(v\ot z^n)&=\pi(f_i)v\ot z^{n-\delta_{i0}},&(actz)\cr
\pi_z(t_i)(v\ot z^n)&=\pi(t_i)v\ot z^n,\cr
\pi_z(q^d)(v\ot z^n)&=v\ot (qz)^n.\cr}
$$
By abuse of notation we sometimes write $V$ for $V_z$ where the context is
clear.

The coproduct $\Delta$ and the antipode $a$ are defined as follows.
$$
\eqalignno{
\Delta(e_i)&=e_i\ot1+t_i\ot e_i,\qquad
\Delta(f_i)=f_i\ot t_i^{-1}+1\ot f_i,&(cop)\cr
\Delta(t_i)&=t_i\ot t_i,\hskip2.05cm \Delta(q^d)=q^d\ot q^d,\cr
a(e_i)&=-t_i^{-1}e_i,\qquad a(f_i)=-f_it_i,
\qquad a(t_i)=t_i^{-1},\qquad a(q^d)=q^{-d}.&(ant)\cr}
$$

\subsection{Fundamental representations and definition of the $R$--matrix}
To each $k~(1\le k\le n)$ we associate a fundamental representation of
$U'_q(\geh)$ denoted by $(\pi^{(k)},V^{(k)})$.
If $k=1$, we call it the vector representation.
If $k=n$ (resp. $k=n-1,n$), we call it the spin representation for $U'_q(\B)$
(resp. $U'_q(\D)$).
The other fundamental representations are
constructed via the fusion construction in the subsequent section.
For each fundamental representation $V^{(k)}$ we shall define linear maps
$\sigma^{(i)}:V^{(k)}\longrightarrow V^{(k')}$ ($i=1$ for $\B$, $i=1,2,3$
for $\D$) satisfying
$$
\eqalignno{
\pi^{(k')}(\sigma^{(i)}(x))\,\sigma^{(i)}(v)&=\sigma^{(i)}(\pi^{(k)}(x)v)
\qquad(x\in U'_q(\geh),v\in V^{(k)}),&(aut)\cr}
$$
where $k'=k$ except that $V^{(n-1)}$ and $V^{(n)}$ may be interchanged
by $\sigma^{(i)}$ for $\geh=\D$.
%***
We use the same notation for the linear maps as for the algebra
automorphisms, in the belief that this should assist the presentation
rather than cause confusion.
We shall consider the maps in detail as the various representations
are presented.
%***

Let $(\pi^{V_i}, V_i)~(i=1,2)$ be arbitrary representations of $U'_q(\geh)$.
Let $\Rb^{V_1V_2}(z)\in\End(V_1\ot V_2)$ be the $R$--matrix, which satisfies
$$
\eqalignno{
&\Rb^{V_1V_2}(z_1/z_2)(\pi^{V_1}_{z_1}\ot\pi^{V_2}_{z_2})\Delta(x)
=(\pi^{V_1}_{z_1}\ot\pi^{V_2}_{z_2})\Delta'(x)\Rb^{V_1V_2}(z_1/z_2),
{}~\forall~x\in U_q(\geh),~~&(defR)\cr}
$$
where $\Delta'=\sigma\circ\Delta$ and $\sigma$ denotes the transposition.
{}From each representation space we normally choose a reference vector
$u_i~(i=1,2)$ and normalise by
$$
\Rb^{V_1V_2}(z)u_1\ot u_2=u_1\ot u_2.
$$
Together with another representation $(\pi^{V_3},V_3)$ of $U'_q(\geh)$, the
$R$-matrices satisfy the Yang--Baxter equation:
$$
\eqalignno{
\Rb^{V_1V_2}(z_1&/z_2)\Rb^{V_1V_3}(z_1/z_3)\Rb^{V_2V_3}(z_2/z_3)\cr
&=\Rb^{V_2V_3}(z_2/z_3)\Rb^{V_1V_3}(z_1/z_3)\Rb^{V_1V_2}(z_1/z_2)\quad
\hbox{on}\quad V_1\ot V_2\ot V_3.&(YBE)\cr}
$$
Here $\Rb^{V_iV_j}(z)$ acts on the third space $V_k$ as the identity.
%***

Let $\scR'(z)$ be the modified universal $R$ defined in appendix 1 of
\cite{IIJMNT92}. Since our $R$--matrix is uniquely determined up to a scalar
from the intertwining property \ref{defR}, we have
$$
(\pi^{V_1}\ot\pi^{V_2})\left(\scR'(z)\right)=\beta^{V_1V_2}(z)\Rb^{V_1V_2}(z).
$$
In order to obtain two point functions by solving the $q$-KZ equation, we
need to know the form of $\beta^{V_1V_2}(z)$. For this purpose,
we consider the second inversion relation
$$
\eqalignno{
&\alpha^{V_1V_2}(z)(((\Rb^{V_1V_2}(z)^{-1})^{t_1})^{-1})^{t_1}
=(q^{-2\ol{\rho}}\ot\id)\Rb^{V_1V_2}(z\xi^{-2})(q^{2\ol{\rho}}\ot\id).
&(2inv)\cr}
$$
Here
%***
$$
\xi=q^{\hc}
$$
and $\alpha^{V_1V_2}(z)$ is a scalar function.
If we replace $\Rb^{V_1V_2}(z)$ by
$(\pi^{V_1}\ot\pi^{V_2})\left(\scR'(z)\right)$ in \ref{2inv} we
know that $\alpha^{V_1V_2}(z)=1$.
As is prescribed in section
4 of \cite{FR92}, this fact enables us to get a difference
equation for $\beta^{V_1V_2}(z)$ and thereby to calculate such factors.

\subsection{Vector representations and their $R$--matrices}
Define an index set $J$ by
$$
\eqalignsp{
J&=\{0,\pm1,\cdots,\pm n\},\phantom{\B}&\hbox{for }\geh=\B,\hskip2cm\cr
 &=\{\pm1,\cdots,\pm n\},\phantom{\D}&\hbox{for }\geh=\D,\hskip2cm\cr}
$$
and set $N=|J|$. We introduce a linear order $\prec$ in $J$ by
$$
\eqalignno{
&1\prec2\prec\cdots\prec n\;(\prec0)\prec-n\prec\cdots\prec-2\prec-1.
\cr}
$$
We shall also use the usual order $<$ in $J$.

We shall define the vector representation $(\pi^{(1)},V^{(1)})$ of
$U'_q(\geh)$.
It is the fundamental representation associated with the vertex 1 in the
Dynkin diagram.
The base vectors of $V^{(1)}$ are given by $\{v_j\mid j\in J\}$.
The weight of $v_j$ is given by $\ep_j(j>0)$, $-\ep_{-j}(j<0)$,
$0\,(j=0)$.  We take $v_1$ as a reference vector.
Set
$$
\eqalignno{
s=&\cases{(-)^n&\quad for $\geh=\B$,\cr
          (-)^{n-1}&\quad for $\geh=\D$.\cr}&(defs)\cr}
$$
Denoting the matrix units by $E_{ij}$ {\it i.e.} $E_{ij}v_k=\delta_{jk}v_i$,
the actions of the generators read as follows
($\pi^{(1)}(f_i)=\pi^{(1)}(e_i)^t$):
$$
\eqalignsp{
\pi^{(1)}(e_0)&=s(E_{-1,2}-E_{-2,1}),\cr
\pi^{(1)}(t_0)&=\sum_{j\in
J}q^{-\delta_{j1}-\delta_{j2}+\delta_{j,-1}+\delta_{j,-2}}
          E_{jj},\cr
\pi^{(1)}(e_i)&=E_{i,i+1}-E_{-i-1,-i}\quad(1\le i\le n-1),\cr
\pi^{(1)}(t_i)&=\sum_{j\in J}q^{\delta_{ji}-\delta_{j,i+1}+\delta_{j,-i-1}
          -\delta_{j,-i}}E_{jj}\quad(1\le i\le n-1),\cr
\pi^{(1)}(e_n)&=\sqrt{[2]_n}(E_{n0}-E_{0,-n})
          &\hbox{for }\geh=\B,\hskip1.5cm\cr
        &=E_{n-1,-n}-E_{n,-n+1}
          &\hbox{for }\geh=\D,\hskip1.5cm\cr
\pi^{(1)}(t_n)&=\sum_{j\in J}q^{\delta_{j,n}-\delta_{j,-n}}E_{jj}
          &\hbox{for }\geh=\B,\hskip1.5cm\cr
        &=\sum_{j\in J}q^{\delta_{j,n-1}+\delta_{jn}-\delta_{j,-n}
          -\delta_{j,-n+1}}E_{jj}
          &\hbox{for }\geh=\D.\hskip1.5cm\cr}
$$

For the vector representation linear maps satisfying \ref{aut} are given as
follows:
$$
\eqalign{
&\sigma^{(i)}:V^{(1)}\longrightarrow V^{(1)},\cr
&\sigma^{(1)}(v_{\pm1})=sv_{\mp1},\cr
&\sigma^{(2)}(v_{\pm n})=v_{\mp n},\cr
&\sigma^{(3)}(v_j)=(-)^{j-1}v_{-(n+1-j)},\cr}
\qquad
\eqalign{
&\phantom{\sigma^{(i)}:V^{(1)}\longrightarrow V^{(1)},}\cr
&\sigma^{(1)}(v_j)=v_j\quad(j\in J,j\ne\pm1),\cr
&\sigma^{(2)}(v_j)=v_j\quad(j\in J,j\ne\pm n),\cr
&\sigma^{(3)}(v_{-j})=(-)^{j-1}sv_{n+1-j}\quad(j>0).\cr}
$$

Let $\{v^*_j\mid j\in J\}$ be the canonical dual basis of $\{v_j\mid j\in J\}$.
Then we have the following isomorphism of $U_q(\geh)$--modules.
$$
\eqalignno{
C^{(1)}_\pm:\quad V^{(1)}_{z\xi^{\mp 1}}
&{\buildrel \sim\over\longrightarrow}\bigl(V^{(1)}_z\bigr)^{*a^{\pm1}};\qquad
v_j\mapsto q^{\overline{\pm j}}v_{-j}^*.
&(dual1)\cr}
$$
Here $\overline{j}$ is defined by
$$
\eqalignno{
\overline{j}&=\cases{j&($j>0$)\cr
                     n&($j=0$)\cr
                     j+N&($j<0$).\cr}
&(jbar)\cr}
$$

Let us recall $R$--matrices for $V^{(1)}\ot V^{(1)}$ obtained in \cite{J86}.
Since the choice of the coproduct is different, we need a slight modification
of the expressions given there.
$\Rb^{(1,1)}(z)$ is given by
$$
\eqalignno{
\Rb^{(1,1)}(z)=&\sum_{i\ne0}E_{ii}\ot E_{ii}
      +{q(1-z)\over1-q^2z}\sum_{i\ne\pm j}E_{ii}\ot E_{jj}\cr
     &+{1-q^2\over1-q^2z}\Bigl(\sum_{i\prec j,i\ne-j}+z\sum_{i\succ j,i\ne-j}
      \Bigr)E_{ij}\ot E_{ji}\cr
     &+{1\over(1-q^2z)(1-\xi z)}\sum_{i,j} a_{ij}(z)E_{ij}\ot E_{-i,-j}.
&(Rvec)\cr}
$$
Here
$$
\eqalignno{
a_{ij}(z)&=\cases{(q^2-\xi z)(1-z)+\delta_{i0}(1-q)(q+z)(1-\xi z)&($i=j$)\cr
         (1-q^2)(q^{\overline{j}-\overline{i}}(z-1)+\delta_{i,-j}(1-\xi z))
               &($i\prec j$)\cr
         (1-q^2)z(\xi q^{\overline{j}-\overline{i}}(z-1)+\delta_{i,-j}(1-\xi
z))
               &($i\succ j$).\cr}
\cr}
$$

For the values of $\alpha^{(1,1)}(z)$ and $\beta^{(1,1)}(z)$ we can obtain
$$
\eqalignno{
\alpha^{(1,1)}(z)&={(1-\xi^{-2}z)(1-q^{-2}\xi^{-1}z)(1-q^2\xi^{-1}z)(1-z)
\over(1-q^2\xi^{-2}z)(1-\xi^{-1}z)^2(1-q^{-2}z)},\cr
\beta^{(1,1)}(z)&=q^{-1}
{\left(q^2z;\xi^2\right)_\infty\left(\xi z;\xi^2\right)_\infty^2
\left(q^{-2}\xi^2z;\xi^2\right)_\infty\over
\left(z;\xi^2\right)_\infty\left(q^{-2}\xi z;\xi^2\right)_\infty
\left(q^2\xi z;\xi^2\right)_\infty
\left(\xi^2z;\xi^2\right)_\infty}.\cr}
$$

\subsection{Spin representations and their $R$--matrices}
We shall define the spin representations. They are the representations
associated with the tail vertices in the Dynkin diagram. There is only one
spin representation in the case of $\B$, and there are two in the case of
$\D$. Consider the 2 dimensional vector space $V_{1/2}$ spanned by
$v_{1/2}$ and $v_{-1/2}$. We define operators $X^+,X^-,T$ acting on this
space by
$$
\eqalignno{
X^{+}v_\eta&=v_{\eta+1},\quad X^{-}v_\eta=v_{\eta-1},\quad
Tv_\eta=q^\eta v_\eta.\cr}
$$
Here $\eta=\pm1/2$, otherwise $v_\eta$ is to be understood as 0.
Set $V^{(sp)}=V_{1/2}^{\ot n}$.
For a base vector
$v_{\vep_1}\ot\cdots\ot v_{\vep_n}~(\vep_1,\cdots,\vep_n=\pm1/2)$ we use
a shorthand notation $v_\vep~(\vep=(\vep_1,\cdots,\vep_n))$.
The weight of $v_\vep$ is given by $\sum_{i=1}^n\vep_i\ep_i$.
We put
$\vep^+=(1/2,\cdots,1/2),\vep^-=(-1/2,\cdots,-1/2)$.
For $i=1,\cdots,n$ and
$\vep=(\vep_1,\cdots,\vep_n)$ $\vep^{(i)}$ (resp. $\vep_{(i)}$) stands for
$(\vep_1,\cdots,{\buildrel i\over {\vep_i+1}},\cdots,\vep_n)$
(resp. $(\vep_1,\cdots,{\buildrel i\over {\vep_i-1}},\cdots,\vep_n)$).
$\vep^{(i)(j)}$, $\vep^{(i)}_{(j)}$, etc, are defined similarly.

We define an action of
$U'_q(\geh)$ on this space as follows ($\pi^{(sp)}(f_i)=\pi^{(sp)}(e_i)^t$):
$$
\eqalignsp{
\pi^{(sp)}(e_0)&=X^{-}\ot X^{-}\ot1\ot\cdots\ot1,\cr
\pi^{(sp)}(t_0)&=T^{-1}\ot T^{-1}\ot1\ot\cdots\ot1,\cr
\pi^{(sp)}(e_i)&=1\ot\cdots\ot{\buildrel i\over {X^+}}
          \ot{\buildrel i+1\over {X^-}}\ot\cdots\ot1&
                  (1\le i\le n-1),\hskip0.5cm\cr
\pi^{(sp)}(t_i)&=1\ot\cdots\ot{\buildrel i\over T}
          \ot{\buildrel i+1\over {T^{-1}}}\ot\cdots\ot1&
                  (1\le i\le n-1),\hskip0.5cm\cr
\pi^{(sp)}(e_n)&=1\ot\cdots\ot1\ot{\buildrel n\over {X^+}}&
                  \hbox{for }\geh=\B,\hskip1cm\cr
        &=1\ot\cdots\ot1\ot{\buildrel n-1\over {X^+}}
          \ot{\buildrel n\over {X^+}}&
                  \hbox{for }\geh=\D,\hskip1cm\cr
\pi^{(sp)}(t_n)&=1\ot\cdots\ot1\ot{\buildrel n\over T}&
                  \hbox{for }\geh=\B,\hskip1cm\cr
        &=1\ot\cdots\ot1\ot{\buildrel n-1\over T}
          \ot{\buildrel n\over T}&
                  \hbox{for }\geh=\D.\hskip1cm\cr}
$$
For this representation linear maps satisfying \ref{aut} are given as follows:
$$
\eqalignno{
&\sigma^{(i)}:V^{(sp)}\longrightarrow V^{(sp)},\cr
&\sigma^{(1)}(v_{(\vep_1\vep_2\cdots\vep_n)})
=v_{(-\vep_1\vep_2\cdots\vep_n)},\cr
&\sigma^{(2)}(v_{(\vep_1\cdots\vep_{n-1}\vep_n)})
%***
=v_{(\vep_1\cdots\vep_{n-1}-\vep_n)},\cr
%***
&\sigma^{(3)}(v_{(\vep_1\cdots\vep_n)})
=v_{(-\vep_n\cdots-\vep_1)}.\cr}
$$
In the case of $\B$ $V^{(sp)}$ is irreducible and is also denoted by
$(\pi^{(n)},V^{(n)})$.
In the case of $\D$ it decomposes into two components.
We denote by $(\pi^{(n)},V^{(n)})$ (resp. $(\pi^{(n-1)},V^{(n-1)})$) the
irreducible representation of $U'_q(\D)$ whose space contains the vector
$v_{\vep^+}$ (resp. $v_{\vep^+_{(n)}}$).
We have
$$
\eqalign{
 \sigma^{(1)},\sigma^{(2)}&:V^{(n)}\longrightarrow V^{(n-1)}\cr
 \sigma^{(3)}&:V^{(n)}\longrightarrow V^{(n-i)}\cr}\qquad
\eqalign{
 \sigma^{(1)},\sigma^{(2)}&:V^{(n-1)}\longrightarrow V^{(n)}\cr
 \sigma^{(3)}&:V^{(n-1)}\longrightarrow V^{(n+i-1)}\cr}
$$
where $i=0$ if $n$ is even and $i=1$ if $n$ is odd.

Denoting by $\{v_\vep^*\}$ the canonical dual basis of $\{v_\vep\}$, we
have the following isomorphism of $U_q(\geh)$--modules.
$$
\eqalignno{
C^{(sp)}_\pm:\quad V^{(sp)}_{z\xi^{\mp 1}}
&{\buildrel \sim\over\longrightarrow}{V^{(sp)}_z}^{*a^{\pm1}};\qquad
v_\vep\mapsto a_\pm(\vep)v_{-\vep}^*,&(dualsp)\cr}
$$
where
$$
\eqalignsp{
a_\pm(\vep)&=(-)^{\sum_{j=1}^n(n+1-j)\ol{\vep}_j}
q^{\pm\sum_{j=1}^n(n+1/2-j)\ol{\vep}_j}&\hbox{for }\geh=\B,\qquad\cr
&=(-q)^{\pm\sum_{j=1}^n(n-j)\ol{\vep}_j}&\hbox{for }\geh=\D,\qquad\cr}
$$
and $\ol{\vep}_j=0$ $(\vep_j=1/2)$, $=1$ $(\vep_j=-1/2)$.
For each irreducible  component in the case of $\D$, $C^{(sp)}_\pm$
alternates the indices of the spaces just like $\sigma^{(3)}$.

We present the explicit forms of the $R$--matrices for $V^{(1)}\ot V^{(n)}$
and $V^{(n)}\ot V^{(1)}$, which we denote by $\Rb^{(1,n)}(z)$ and
$\Rb^{(n,1)}(z)$. We take $v_{\vep^+}$ in $V^{(n)}$ as a reference vector
for the both cases. For the case of $\D$ we may also need the $R$--matrices
for $V^{(1)}\ot V^{(n-1)}$ and $V^{(n-1)}\ot V^{(1)}$. Those are given by
the same formulae as $\Rb^{(1,n)}(z)$ and $\Rb^{(n,1)}(z)$ except that they
are acting on different spaces $V^{(1)}\ot V^{(n-1)}$ and
$V^{(n-1)}\ot V^{(1)}$. As a reference vector for $V^{(n-1)}$ in the case of
$\D$ we take $v_{\vep^+_{(n)}}$.

For $i\in J$ we set $\vep(i)=\vep^{(i)}(i>0)$, $=\vep(i=0)$,
$=\vep_{(-i)}(i<0)$.
$\vep(i,j)~(i,j\in J)$ is defined recursively.
For $i,j=1,\cdots,n~(i<j)$ define
$n^+_{ij}(\vep)=\sharp\{k\mid i<k<j,\vep_k=1/2\}$,
$n^+_{i0}(\vep)=\sharp\{k\mid i<k,\vep_k=1/2\}$ and put
$n^+_{ji}(\vep)=-n^+_{ij}(\vep)-1$ for $i,j=0,\cdots,n$.
The forms of the $R$--matrices are given below.
$$
\eqalignno{
\Rb^{(1,n)}(z)&=\sum_{i,\vep}a_{i\vep}(z)E_{ii}\ot E_{\vep\vep}
               +\sum_{i\neq j,\vep}b_{ij\vep}c_{ij}(z)
               (\xi^{1/2}sz)^{\th(i\succ j)}E_{ij}\ot E_{\vep\vep(i,-j)},\cr
\Rb^{(n,1)}(z)&=\sum_{i,\vep}a_{i\vep}(z)E_{\vep\vep}\ot E_{ii}
               +\sum_{i\neq j,\vep}\ol{b}_{ij\vep}c_{ij}(z)
               (\xi^{1/2}sz)^{\th(i\prec j)}E_{\vep\vep(i,-j)}\ot E_{ij},\cr}
$$
where
$$
\eqalignno{
a_{i\vep}(z)&=\cases{1&($i\vep_{|i|}>0$),\cr
  q\displaystyle{1+q^{-1}\xi^{1/2}sz\over
    1+q\xi^{1/2}sz}&($i\vep_{|i|}<0$),\cr
  q^{1/2}\displaystyle{1+\xi^{1/2}sz\over
    1+q\xi^{1/2}sz}&($i=0$),\cr}\cr
b_{ij\vep}&=q^{-\eta(i)+\eta(j)+\delta_{i0}}(-q)^{n^+_{|i||j|}(\vep)},\cr
\ol{b}_{ij\vep}&=q^{\eta(i)-\eta(j)+(\delta_{i0}+\delta_{j0})/2}
                (-q)^{n^+_{|j||i|}(\vep)},\cr
\eta(i)&=\cases{0&($i\geq0$),\cr \hc/2+i+1&($i<0$),\cr}\cr
c_{ij}(z)&=\cases{
  \displaystyle{1-q^2\over
   1+q\xi^{1/2}sz}&($ij\neq0$),\cr
  \sqrt{[2]_n}\displaystyle{1-q\over1+q\xi^{1/2}sz}&($ij=0$).\cr}\cr}
$$
Here $\th(S)=1$ if $S$ is true, $=0$ otherwise.
In the summation $i,j$ run over $J$ and $\vep$ runs over $(\pm1/2,\cdots,
\pm1/2)$. $E_{\vep\vep'}$ stands for a matrix unit in $\End(V^{(n)})$.
The calculation is long but straightforward. We omit it.

For these $R$--matrices we have
$$
\eqalignno{
\alpha^{(1,n)}(z)&=\alpha^{(n,1)}(z)
       ={(1+q\xi^{-1/2}sz)(1+q^{-1}\xi^{-3/2}sz)\over
         (1+q^{-1}\xi^{-1/2}sz)(1+q\xi^{-3/2}sz)},\cr
\beta^{(1,n)}(z)&=\beta^{(n,1)}(z)
=q^{-1/2}{(-q\xi^{1/2}sz;\xi^2)_\infty(-q^{-1}\xi^{3/2}sz;\xi^2)_\infty\over
          (-q^{-1}\xi^{1/2}sz;\xi^2)_\infty(-q\xi^{3/2}sz;\xi^2)_\infty}.\cr}
$$
%
%
%               Section 4
%
%
\section{Fusion construction}

Throughout this section, $(\pi,V)$ stands for the vector representation of
$\up$  defined in section 3, and $P$ denotes the transposition on $V\ot V$.
We shall construct the fundamental representations $(\pi^{(k)},V^{(k)})$
of $\up$ via the fusion construction, where $2\le k\le n-1$ for $\geh=\B$
and $2\le k\le n-2$ for $\geh=\D$.

\subsection{Definition of $V^{(k)}$}
Since $\Rb^{(1,1)}(z)$ has a pole at $z=q^{-2}$, we set
$$
\eqalignno{
R(z)&=(1-q^2z)\Rb^{(1,1)}(z).&(rdefR)\cr}
$$
Let $R_{ii+1}(z)$ be an operator on $V^{\ot k}$ acting as $R(z)$ on the
$i$-th and $(i+1)$-th components and as the identity on the other components,
and let
$$
\eqalignno{
S_{ij}&=R_{j-1j}(q^{-2})R_{j-2j}(q^{-4})\cdots R_{ij}(q^{2i-2j})\quad(i<j),\cr
T^{(k)}&=S_{1k}\cdots S_{13}S_{12}.&(defT)\cr}
$$
We define $V^{(k)}$ by
$$
\eqalignno{
V^{(k)}&=V^{\ot k}\big/\Ker T^{(k)}.\cr}
$$
%***
A graphical representation of $T^{(4)}$ is given in Figure 3.

\midinsert
\vbox{\vskip 0.8cm
\centerline{\epsfxsize=5cm\epsfbox{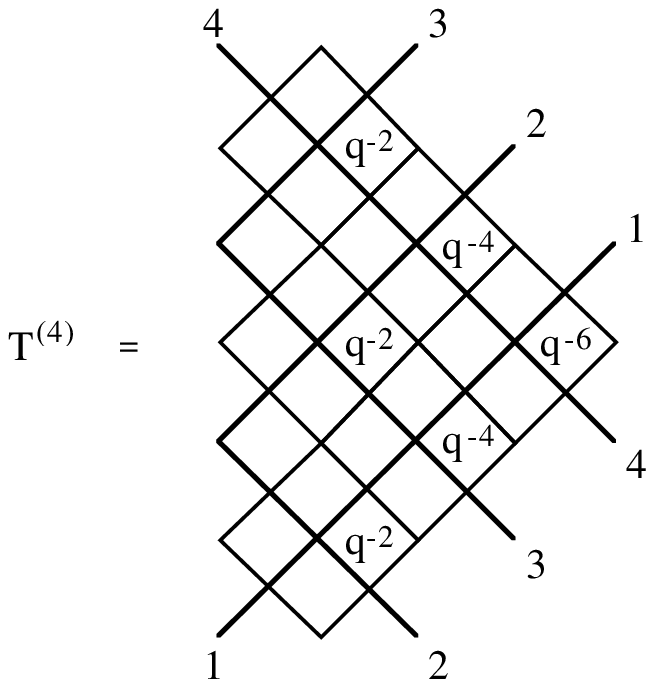}}
\vskip 0.5cm
\centerline{{\bf Figure 3:} Graphical representation of $T^{(4)}$}
\vskip 0.5cm}
\endinsert

We shall give a more concrete description of the space $V^{(k)}$. Set
$W=\Im PR(q^2)$ $\subset V^{\ot2}$.

\prop{defW}.
$W$ is generated by the following vectors.
$$
\eqalignno{
&v_i\ot v_i\quad(i\neq0),\qquad\qquad v_i\ot v_j+qv_j\ot v_i
\quad(i\succ j,~i\not=-j),\cr
&v_{-i}\ot v_i+q^2v_i\ot v_{-i}-q(v_{i+1}\ot v_{-i-1}+v_{-i-1}\ot v_{i+1})
\quad(1\leq i\leq n-1),\cr
&v_{-n}\ot v_n+q^2v_n\ot v_{-n}-q(1+q)(v_0\ot v_0)
\qquad\hbox{for }\geh=\B.\cr}
$$
\endprop
The proof is given in the appendix. $\Ker T^{(k)}$ has the following concrete
description.

\prop{Ker}.
$$
\Ker T^{(k)}=\sum_{j=0}^{k-2}V^{\ot j}\ot W\ot V^{\ot(k-2-j)},
$$
\endprop

Before proving this proposition we state a few properties of $V^{(k)}$.
By virtue of these propositions it is easily seen that the vectors of the
following form constitute a basis of $V^{(k)}$:
$$
\eqalign{
&v_{i_1}\ot\cdots v_{i_a}\ot v_0\ot\cdots\ot v_0\ot
v_{-j_b}\ot\cdots\ot v_{-j_1}\cr
&v_{i_1}\ot\cdots v_{i_a}\ot v_{-n}\ot v_n\ot v_{-n}\ot\cdots\ot v_n\ot
v_{-j_b}\ot\cdots\ot v_{-j_1}\cr}
{}~\eqalign{
&\hbox{for }\geh=\B,\cr
&\hbox{for }\geh=\D.\cr}
\eqno(nord)
$$
%*** from here to next %***
Here $1\le i_1<\cdots<i_a\le n,1\le j_1<\cdots<j_b\le n$. We shall call such
a form the {\it normal order} form. Using this basis we easily get
$$
\dim V^{(k)}
=\rank T^{(k)}
=\sum_{j=0}^{[k/2]}{N\choose k-2j}.
$$
Here $N=\dim V$ and $[i]$ denotes the largest integer that does not
exceed $i$.
%The following formula is sometimes useful.
%$$
%\eqalignno{
%v_{-i}\ot v_i&\equiv-q^2v_i\ot v_{-i}+(1-q^2)
%\sum_{j=1}^{n-i}q^jv_{i+j}\ot v_{-i-j}\cr
%&\qquad+\cases{q^{n-i+1}(1+q)v_0\ot v_0&($\geh=\B$)\cr
%               q^{n-i}v_{-n}\ot v_n&($\geh=\D$)\cr}\quad\mod W.\cr}
%$$

\Proofof{\refProp{Ker}}
By successive use of the Yang--Baxter equation \ref{YBE} for $R(z)$, for
any $j~(0\le j\le k-2)$, we can rewrite $T^{(k)}$ into the following form:
$$
T^{(k)}=T'_jR_{j+1j+2}(q^{-2})\qquad
\hbox{with some operator $T'_j$ on $V^{\ot k}$}.
$$
Note that $R(q^{-2})PR(q^2)=0$. So $T^{(k)}$ kills a vector in $V^{\ot j}
\ot W\ot V^{\ot(k-2-j)}$ for any $j~(0\le j\le k-2)$. This shows
$\Ker T^{(k)}\supset\sum_{j=0}^{k-2}V^{\ot j}\ot W\ot V^{\ot(k-2-j)}$.
To prove the other inclusion it suffices to show
%$$
%\dim \Ker T^{(k)}\le \dim \sum_{j=0}^{k-2}V^{\ot j}\ot W\ot V^{\ot(k-2-j)},
%$$
%or equivalently,
$\rank T^{(k)}\ge \sum_{j=0}^{[k/2]}{N\choose k-2j}$.
Put
$$
T^{\br{k}}=\lim_{q\to1}{T^{(k)}\over(1-q^2)^{k(k-1)/2}}.
$$
To conclude it suffices to show
$$
\eqalignno{
\rank T^{\br{k}}&\ge \sum_{j=0}^{[k/2]}{N\choose k-2j},&(rk)\cr}
$$
which will be shown in the appendix.
\qed

\subsection{Action of $\uq$}
We define the action of $\uq$ on $V^{(k)}_z$ below. Define
$$
\eqalignno{
\pi^{(k)}_z&:\uq\longrightarrow\End(V^{\ot k})\cr
\pi^{(k)}_z(x)&=(\pi_{(-q)^{1-k}z}\ot\pi_{(-q)^{3-k}z}\ot\cdots\ot
\pi_{(-q)^{k-1}z})\Delta^{(k)}(x),\cr}
$$
where
$$
\Delta^{(k)}=(\Delta\ot\underbrace{\id\ot\cdots\ot\id}_{k-2})
\circ\Delta^{(k-1)},\qquad\Delta^{(2)}=\Delta.
$$
In what follows we use running index notation.
$$
\Delta^{(k)}(x)=\sum x_{(1)}\ot\cdots\ot x_{(k)}.
$$

\prop{act}.
$\pi^{(k)}_z(x)$ is well defined as an operator on $V^{(k)}_z$, that is,
$$
\pi^{(k)}_z(x)\Ker T^{(k)}\subset \Ker T^{(k)}.
$$
\endprop

\Proof
Put
$$
\pi^{(k)\prime}_z(x)=(\pi_{(-q)^{1-k}z}\ot\pi_{(-q)^{3-k}z}\ot\cdots\ot
\pi_{(-q)^{k-1}z})\sum x_{(k)}\ot\cdots\ot x_{(1)}.
$$
Using the intertwining property of $R(z)$ \ref{defR} we can show
$$
T^{(k)}\pi^{(k)}_z(x)=\pi^{(k)\prime}_z(x)T^{(k)}.
$$
This completes the proof.
\qed

As a $U_q(\ol{{\geh}})$--module we have the following isomorphism:
$$
\eqalignno{
V^{(k)}&\cong V_{\ol{\La}_k}\oplus V_{\ol{\La}_{k-2}}\oplus\cdots\oplus
(V_{\ol{\La}_1}\hbox{ or }V_0),&(decomp)\cr}
$$
where $V_{\ol{\La}_j}$ is the irreducible highest weight module with highest
weight $\ol{\La}_j$ and $V_0$ is the trivial module.

%*** the following is changed
Next we consider the Dynkin diagram symmetries of the $\up$--module $V^{(k)}$:
$$
\sigma^{(i)}:V^{(k)}\longrightarrow V^{(k)}.
$$
We shall give the action of $\sigma^{(i)}~(i=1,2,3)$, by components, {\it i.e.}
$$
\eqalignno{
\sigma^{(i)}(v_{i_1}\ot\cdots\ot v_{i_k})
&=\sigma^{(i)}_1(v_{i_1})\ot\cdots\ot\sigma^{(i)}_k(v_{i_k}),\cr}
$$
where $\sigma^{(i)}_m$'s are given by the following rules.
$$
\eqalignno{
\sigma^{(1)}_m(v_{\pm1})&=(-q)^{\pm(2m-k-1)}sv_{\mp1},\cr
\sigma^{(1)}_m(v_j)&=v_j\hskip4.1cm(j\in J,~j\neq\pm1),\cr
\sigma^{(2)}_m(v_{\pm n})&=v_{\mp n},\cr
\sigma^{(2)}_m(v_j)&=v_j\hskip4.1cm(j\in J,~j\neq\pm n),\cr
\sigma^{(3)}_m(v_j)&=(-q)^{2m-k-1\over2}(-)^{j-1}v_{-(n+1-j)},\cr
\sigma^{(3)}_m(v_{-j})&=(-q)^{k+1-2m\over2}(-)^{j-1}sv_{n+1-j}\quad(j>0).\cr}
$$
%***

Extending \ref{dual1} we have an isomorphism of $\uq$--modules
$$
\eqalignno{
C^{(k)}_\pm:\quad V^{(k)}_{z\xi^{\mp 1}}
&{\buildrel \sim\over\longrightarrow}\bigl(V^{(k)}_z\bigr)^{*a^{\pm1}}.
&(dualk)\cr}
$$
The proof is given in the appendix.

\subsection{Definition of ${\ol{R}^{(1,k)}(z)}$ and ${\ol{R}^{(k,1)}(z)}$}
We define operators
$$
\eqalignno{
R^{(1,k)}(z)&\in\End({\buildrel 0\over V}\ot{\buildrel 1\over V}\ot\cdots\ot
{\buildrel k\over V}),\cr
R^{(k,1)}(z)&\in\End({\buildrel 1\over V}\ot\cdots\ot{\buildrel k\over V}\ot
{\buildrel k+1\over V})\cr}
$$
by
$$
\eqalignno{
R^{(1,k)}(z)&=R_{0k}((-q)^{-k+1}z)\cdots R_{02}((-q)^{k-3}z)
R_{01}((-q)^{k-1}z),\cr
R^{(k,1)}(z)&=R_{1k+1}((-q)^{-k+1}z)\cdots R_{k-1k}((-q)^{k-3}z)
R_{kk+1}((-q)^{k-1}z).
\cr}
$$

\prop{wdefR}.
$R^{(1,k)}(z)$ (resp. $R^{(k,1)}(z)$) is well defined as an operator on
$V\ot V^{(k)}$ (resp. $V^{(k)}\ot V$), that is,
$$
\eqalignno{
R^{(1,k)}(z)(V\ot\Ker T^{(k)})&\subset V\ot\Ker T^{(k)},\cr
R^{(k,1)}(z)(\Ker T^{(k)}\ot V)&\subset \Ker T^{(k)}\ot V.\cr}
$$
\endprop

\Proof
These can be proved from the following formulae:
$$
\eqalignno{
({\id}_V\ot T^{(k)})R^{(1,k)}(z)
&=R_{01}((-q)^{k-1}z)\cdots R_{0k}((-q)^{-k+1}z)({\id}_V\ot T^{(k)}),\cr
(T^{(k)}\ot{\id}_V)R^{(k,1)}(z)
&=R_{kk+1}((-q)^{k-1}z)\cdots R_{1k+1}((-q)^{-k+1}z)(T^{(k)}\ot{\id}_V),\cr}
$$
which are shown easily by successive use of the Yang--Baxter equation
\ref{YBE}.
\qed

We shall show that $R^{(1,k)}$ and $R^{(k,1)}$ have the intertwining property
in the sense of \ref{defR}.

\prop{int}.
$$
\eqalignno{
R^{(1,k)}(z_1/z_2)(\pi_{z_1}\ot\pi^{(k)}_{z_2})\Delta(x)&=
(\pi_{z_1}\ot\pi^{(k)}_{z_2})\Delta'(x)R^{(1,k)}(z_1/z_2)
\quad\hbox{on }V\ot V^{(k)},\cr
R^{(k,1)}(z_1/z_2)(\pi^{(k)}_{z_1}\ot\pi_{z_2})\Delta(x)&=
(\pi^{(k)}_{z_1}\ot\pi_{z_2})\Delta'(x)R^{(k,1)}(z_1/z_2)
\quad\hbox{on }V^{(k)}\ot V.\cr}
$$
\endprop

\Proof
The upper formula can be shown in the following manner.
$$
\eqalignno{
R^{(1,k)}&(z)
\sum\pi_{z_1}(x_{(1)})\ot\pi^{(k)}_{z_2}(x_{(2)})\cr
&=R_{0k}((-q)^{-k+1}z)\cdots R_{01}((-q)^{k-1}z)\cr
&\quad
\sum\pi_{z_1}(x_{(1)})\ot\pi_{(-q)^{1-k}z_2}(x_{(2)})\ot\cdots\ot
\pi_{(-q)^{k-1}z_2}(x_{(k+1)})\cr
&=R_{0k}((-q)^{-k+1}z)\cdots R_{02}((-q)^{k-3}z)\cr
&\quad
\sum\pi_{z_1}(x_{(2)})\ot\pi_{(-q)^{1-k}z_2}(x_{(1)})\ot
%\pi_{(-q)^{3-k}z_2}(x_{(3)})\ot
\cdots\ot\pi_{(-q)^{k-1}z_2}(x_{(k+1)})R_{01}((-q)^{k-1}z)\cr
&~~\vdots\cr
&=
\sum\pi_{z_1}(x_{(k+1)})\ot\pi_{(-q)^{1-k}z_2}(x_{(1)})\ot\cdots\ot
\pi_{(-q)^{k-1}z_2}(x_{(k)})R^{(1,k)}(z)\cr
&=
\sum\pi_{z_1}(x_{(2)})\ot\pi^{(k)}_{z_2}(x_{(1)})R^{(1,k)}(z).\cr}
$$
The other formula can be shown similarly.
\qed

As a reference vector for $V^{(k)}$ we take $v_{[12\cdots k]}$, where
$v_{[i_1\cdots i_k]}=v_{i_1}\ot\cdots\ot v_{i_k}$. We want to normalise
$R^{(1,k)}(z)$ (resp. $R^{(k,1)}(z)$) so that
$$
\eqalignno{
R^{(1,k)}(z)v_1\ot v_{[12\cdots k]}&=v_1\ot v_{[12\cdots k]}\cr
(\hbox{resp. }R^{(k,1)}(z)v_{[12\cdots k]}\ot v_1
&=v_{[12\cdots k]}\ot v_1).\cr}
$$

\prop{nor}.
$$
\eqalignno{
\Rb^{(1,k)}(z)&={1\over q^{k-1}(1-(-q)^{k+1}z)
\prod_{j=1}^{k-1}(1-(-q)^{k-1-2j}z)}
R^{(1,k)}(z),\cr
\Rb^{(k,1)}(z)&={1\over q^{k-1}(1-(-q)^{k+1}z)
\prod_{j=1}^{k-1}(1-(-q)^{k-1-2j}z)}
R^{(k,1)}(z).\cr}
$$
\endprop

\Proof
$R^{(1,k)}(z)v_1\ot v_{[12\cdots k]}$ produces many types of vectors
in  $V^{\ot k}$.
One such vector is obtained by applying each
$R_{0j}((-q)^{k+1-2j}z)$ diagonally.
It is
$$
\eqalignno{
R^{(1,k)}&(z)(v_1\ot v_1\ot v_2\ot\cdots\ot v_k)\cr
&=(1-(-q)^{k+1}z)\cdot q(1-(-q)^{k-3}z)\cdot q(1-(-q)^{k-5}z)\cr
&\qquad\qquad\cdots q(1-(-q)^{-k+1}z)
(v_1\ot v_1\ot v_2\ot\cdots\ot v_k).\cr}
$$
All of the off-diagonal terms are zero in the quotient space $V^{(k)}$
due to the definition of $R^{(1,k)}(z)$ as a product $R_{0k}\cdots
R_{01}$. For example, $R_{02}(v_1\ot v_{[12\cdots k]})$ produces two
terms, the diagonal one above and $(v_2\ot v_{[11\cdots k]})$.
The latter term is zero in $V^{(k)}$.
The argument proceeds step by step.
The other case can be shown similarly.
\qed

\subsection{Image of the universal $R$}
Proceeding exactly as for the vector and spin representations, we find
$$
\eqalignno{
&\alpha^{(1,k)}(z)=\alpha^{(k,1)}(z)\cr
&\quad={(1-(-q)^{k-1}\xi^{-2}z)(1-(-q)^{-k-1}\xi^{-1}z)
(1-(-q)^{k+1}\xi^{-1}z)(1-(-q)^{-k+1}z)
\over
(1-(-q)^{k+1}\xi^{-2}z)(1-(-q)^{-k+1}\xi^{-1}z)
(1-(-q)^{k-1}\xi^{-1}z)(1-(-q)^{-k-1}z)},\cr
&\beta^{(1,k)}(z)=\beta^{(k,1)}(z)\cr
&\quad=q^{-1}
{((-q)^{k+1}z;\xi^2)_\infty((-q)^{-k+1}\xi z;\xi^2)_\infty
((-q)^{k-1}\xi z;\xi^2)_\infty((-q)^{-k-1}\xi^2 z;\xi^2)_\infty\over
((-q)^{k-1}z;\xi^2)_\infty((-q)^{-k-1}\xi z;\xi^2)_\infty
((-q)^{k+1}\xi z;\xi^2)_\infty((-q)^{-k+1}\xi^2 z;\xi^2)_\infty}.\cr}
$$
%
%
%               Section 5
%
%
\section{Vertex operators}
In this section we shall define vertex operators (\VO s), and review necessary
properties. We also fix the normalisation of each \VO\ needed in subsequent
sections.

\subsection{Definition of \VO s}
Let $V(\La_i)$ be the irreducible highest weight module with highest weight
$\La_i$. Let $\ket{\La_i}$ be a highest weight vector of $V(\La_i)$. As
motivated in section 2 we consider level 1 modules only, {\it i.e.} $i=0,1,n$
(and $n-1$ for $\geh=\D$).
Let $\la,\mu$ stand for level 1 weights.
We define
two types of \VO s as intertwiners of the following $\uq$--modules.
$$
\eqalignno{
\Phit_\la^{\mu V^{(k)}}(z)&:\quad
V(\la)\longrightarrow V(\mu)\ot V^{(k)}_z,&(vo:a)\cr
\Phit_\la^{V^{(k)}\mu}(z)&:\quad
V(\la)\longrightarrow V^{(k)}_z\ot V(\mu).&(vo:b)\cr}
$$
Set $\Delta_\la=(\la|\la+2\rho)/2(\hc+1)$.
We shift the powers of $z$ by these `conformal factors':
$$
\Phi_\la^{\mu V^{(k)}}(z)
=z^{\Delta_\mu-\Delta_\la}\Phit_\la^{\mu V^{(k)}}(z),\qquad
\Phi_\la^{V^{(k)}\mu}(z)
=z^{\Delta_\mu-\Delta_\la}\Phit_\la^{V^{(k)}\mu}(z),
$$
and call them type I \VO s and type II \VO s respectively.
For a \VO\ \ref{vo:a} we  define its `leading term' $v_{lt}$ by
$$
\Phit_\la^{\mu V^{(k)}}(z)\ket{\la}=\ket{\mu}\ot v_{lt}+\cdots.
$$
Here the terms in $\cdots$ should be of the form $\ket{u}\ot v$ with
$\ket{u}\in\bigoplus_{\xi\ne\mu}V(\mu)_\xi$.
Set
$$
\bigl(V^{(k)}\bigr)_\la^\mu=\{v\in V^{(k)}\mid \la\equiv\mu+\wt v \mod \delta,
{}~e_i^{\langle h_i,\mu\rangle+1}v=0\hbox{ for all }i\},
$$
The following criterion for the existence of \VO s is known \cite{DJO93}.
\prop{crit}.
Mapping a \VO\ to its leading term gives an isomorphism of vector spaces.
$$
\{\VO s:V(\la)\rightarrow V(\mu)\ot V^{(k)}_z\}
{\buildrel \sim\over\longrightarrow} \bigl(V^{(k)}\bigr)_\la^\mu
$$
\endprop
The criterion for the existence of type II \VO s is similar.

\subsection{Two point functions}
Here we collect useful propositions to calculate needed two point functions.
Most of these come from appendix 1 of \cite{IIJMNT92}. We consider the
following three types of two point functions.
$$
\eqalignno{
\br{\nu|&\Phi_\mu^{\nu W_2}(z_2)\Phi_\la^{\mu V_1}(z_1)|\la}&(tpf:a)\cr
\br{\nu|&\Phi_\mu^{W_2 \nu}(z_2)\Phi_\la^{\mu V_1}(z_1)|\la}&(tpf:b)\cr
\br{\nu|&\Phi_\mu^{\nu W_2}(z_2)\Phi_\la^{V_1 \mu}(z_1)|\la}&(tpf:c)\cr}
$$
Following \cite{IIJMNT92}, the space indexed by 1 (resp.$\,$2) should always
come in the first (resp. second) component of the tensor product. For
instance, \ref{tpf:a} stands for the expectation value of the following
composition of \VO s.
$$
V(\la)
{\buildrel \Phi_\la^{\mu V}(z_1)\over\longrightarrow}
V(\mu)\ot V_{z_1}
{\buildrel \Phi_\mu^{\nu W}(z_1)\ot1\over\longrightarrow}
V(\nu)\ot W_{z_2}\ot V_{z_1}
{\buildrel 1\ot P\over\longrightarrow}
V(\nu)\ot V_{z_1}\ot W_{z_2}.
$$
\ref{tpf:b} is similar, but in \ref{tpf:c} we don't need the last
transposition. We call a two point function of type \ref{tpf:a},
\ref{tpf:b},\ref{tpf:c} the type (I,I),(I,II),(II,I) two point
function, respectively.

For finite dimensional representations $\pi^V:\up\rightarrow\End(V)$,
$\pi^W:\up\rightarrow\End(W)$, we put
$$
R_+^{VW}(z)=(\pi^V\ot\pi^W)(\scR'(z)).
$$
%***
Then the two point functions are determined as solutions of the $q$--KZ
equation (with appropriate analyticity properties):
\prop{qKZ}~\hbox{\rm\cite{FR92,IIJMNT92}}.
Let $\Psi(z_1,z_2)$ be one of \ref{tpf:{a-c}}. Then we have
$$
\Psi(pz_1,z_2)=A(z_1/z_2)\Psi(z_1,z_2),\quad
\Psi(pz_1,pz_2)=(q^{-\phi}\ot q^{-\phi})\Psi(z_1,z_2),
$$
where $p=q^{2(\hc+1)},\phi={\bar \la}+{\bar \nu}+2{\bar \rho}$ and
$$
\eqalignsp{
A(z)&=R_+^{VW}(pz)(q^{-\phi}\ot1)
    &\hbox{\rm for \ref{tpf:a}}\hskip.5cm\cr
&=(q^{-{\bar \nu}}\ot1)R_+^{VW}(pq^{-1}z)(q^{-\phi+{\bar \nu}}\ot1)
    &\hbox{\rm for \ref{tpf:b}}\hskip.5cm\cr
&=(q^{-\phi+{\bar \nu}}\ot1)R_+^{VW}(qz)(q^{-{\bar \nu}}\ot1)
    &\hbox{\rm for \ref{tpf:c}}\hskip.5cm\cr}
$$
\endprop

For our calculation the following propositions are quite useful.

\prop{usefulprop1}~\hbox{\rm\cite{IIJMNT92}}.
\item{(i)}Consider $\Psi(z_1,z_2)$ in the case \ref{tpf:a}. Then for any $i$
we have
$$
(\pi^V_{z_1}\ot\pi^W_{z_2})\Delta'(e_i)^{\br{h_i,\nu}+1}\Psi(z_1,z_2)=0,
\qquad \wt \Psi(z_1,z_2)={\bar \la}-{\bar \nu}.
$$
\item{(ii)}Let $\Psi(z_1,z_2)$ be a solution for \ref{tpf:a}. Then the
following give solutions for the other cases:
$$
\eqalignsp{
&(q^{-{\bar \nu}}\ot1)\Psi(q^{-1}z_1,z_2)
   &\hbox{\rm for \ref{tpf:b}}\hskip.5cm\cr
&(q^{-\phi+{\bar \nu}}\ot1)\Psi(p^{-1}qz_1,z_2)
   &\hbox{\rm for \ref{tpf:c}}\hskip.5cm\cr}
$$
\endprop
\prop{usefulprop2}.
Assume that for a $V\ot W$--valued function $w(z)$ we have
$$
\eqalignno{
&(\pi^V_{z_1}\ot\pi^W_{z_2})\Delta'(e_i)^{\br{h_i,\nu}+1}w(z_1/z_2)=0
\qquad\hbox{for all $i$},&(asp1)\cr
&\Rb^{VW}(pz)(q^{-\phi}\ot1)w(z)=r(z)w(pz),&(asp2)\cr}
$$
with some scalar function $r(z)$.
Then setting $\ol{w}(z)=P(q^{-\phi}\ot1)w(p^{-1}z^{-1})$, we get
$$
\eqalignno{
&(\pi^W_{z_1}\ot\pi^V_{z_2})\Delta'(e_i)^{\br{h_i,\nu}+1}\ol{w}(z_1/z_2)=0
\qquad\hbox{for all $i$},\cr
&\Rb^{WV}(pz)(q^{-\phi}\ot1)\ol{w}(z)
=q^{-\br{\phi,{\rm wt}\ol{w}}}r(p^{-2}z^{-1})^{-1}\ol{w}(pz).\cr}
$$
\endprop

\Proof
{}From \ref{asp1}, for any $i$ we have
$$
\Rb^{WV}(z^{-1})P(\pi^V_{z_1}\ot\pi^W_{z_2})
\Delta'(e_i)^{\br{h_i,\nu}+1}w(z_1/z_2)=0.
$$
The LHS is equal to
$$
(\pi^W_{z_2}\ot\pi^V_{z_1})\Delta'(e_i)^{\br{h_i,\nu}+1}
\Rb^{WV}(z^{-1})Pw(z).
$$
Here using the first inversion relation and \ref{asp2} we can get
$$
\eqalignno{
\Rb^{WV}(z)Pw(z^{-1})&=P\Rb^{VW}(z^{-1})^{-1}w(z^{-1})\cr
&=r(p^{-1}z^{-1})^{-1}P(q^{-\phi}\ot1)w(p^{-1}z^{-1}).\cr}
$$
This shows the first equality. The second can also be shown using the same
formulae.
\qed

\subsection{Normalisation of \VO s}
%*** the next paragraph has been extensively re-written
%***
In this subsection we shall fix the normalisations of \VO s. We start with
recalling Dynkin diagram automorphisms \ref{Dda}. For level 1 highest
weight modules there exist linear maps
$$
\eqalign{
&\sigma^{(1)}:\quad V(\La_i)\longrightarrow V(\La_{1-i})\cr
&\sigma^{(2)}:\quad V(\La_{n-i})\longrightarrow V(\La_{n+i-1})\cr
&\sigma^{(3)}:\quad V(\La_i)\longrightarrow V(\La_{n-i})\cr}
\qquad\eqalign{
&(i=0,1),\cr
&(i=0,1),\cr
&(i=0,1,n-1,n),\cr}
\eqno(authwm)
$$
such that, for arbitrary $x\in\up$, $\ket{u}\in V(\la)$,
$\sigma^{(i)}\ket{u}\in V(\sigma^{(i)}(\la))$,
$\sigma^{(i)}$ maps the fixed highest weight vector $\ket{\la}$ to
the highest weight vector $\ket{\sigma^{(i)}(\la)}$ and preserves the algebra
action: $\sigma^{(i)}(x\ket{u})=\sigma^{(i)}(x)\sigma^{(i)}(\ket{u})$.
(For a level 1 weight $\la$, $\sigma(\la)$ stands for the corresponding
weight by \ref{authwm}.)
Note that we have again used the same notation as \ref{Dda}.
{}From the Dynkin diagram symmetry,
if there exists a unique intertwiner $V(\la)\longrightarrow V(\mu)\ot
V^{(k)}_z$ (up to a scalar factor), then there is also such an intertwiner
$\sigma(V(\la))\longrightarrow\sigma(V(\mu))\ot\sigma(V^{(k)})_z$,
which may be obtained by composition of the linear maps $\sigma^{(i)}$
($i=1,2,3$).
The condition for uniqueness is
$$
\eqalignno{
&\dim(V^{(k)})_\la^\mu=1,\cr}
$$
and the \VO s appearing herein have this property.
If we choose the normalisation
$$
\eqalignno{
&\Phit_\la^{\mu V^{(k)}}(z)\ket{\la}=\ket{\mu}\ot v_{lt}+\cdots
\qquad(v_{lt}:\hbox{ leading term}),\cr}
$$
then we can use uniqueness to fix the normalisation of
$\Phit_{\sigma(\la)}^{\sigma(\mu)\sigma(V^{(k)})}(z)$ by
$$
\eqalignno{
&\Phit_{\sigma(\la)}^{\sigma(\mu)\sigma(V^{(k)})}(z)\sigma(\ket{\la})
=\sigma(\ket{\mu})\ot \sigma(v_{lt})+\cdots.\cr}
$$
We shall use this extensively to minimise calculations.
%***

For the vector representation $V^{(1)}$ we fix the normalisations as
follows:
$$
\eqalign{
\Phit_{\La_1}^{\La_0 V^{(1)}}(z)\ket{\La_1}
&=\ket{\La_0}\ot v_1+\cdots\cr
\Phit_{\La_n}^{\La_n V^{(1)}}(z)\ket{\La_n}
&=\ket{\La_n}\ot \alpha^{-1}v_0+\cdots\cr}
\qquad\eqalign{
&\hbox{for }\geh=\B,\D,\cr
&\hbox{for }\geh=\B,\cr}
$$
where
$$
\alpha=\sqrt{[2]_n}.
\eqno(defalpha)
$$
The \VO s $\Phit_{\La_0}^{\La_1 V^{(1)}}$ ($\geh=\B,\D$) and
$\Phit_{\La_{n-1}}^{\La_n V^{(1)}}$,  $\Phit_{\La_n}^{\La_{n-1}V^{(1)}}$
($\geh=\D$) are normalised using  Dynkin diagram automorphisms.

For the spin representations we normalise by
$$
\eqalign{
\Phit_{\La_n}^{\La_0 V^{(n)}}(z)\ket{\La_n}
&=\ket{\La_0}\ot v_{\vep^+}+\cdots\cr
\Phit_{\La_0}^{\La_n V^{(n)}}(z)\ket{\La_0}
&=\ket{\La_n}\ot v_{\vep^-}+\cdots\cr}
\qquad\eqalign{
&\hbox{for }\geh=\B,\D,\cr
&\hbox{for }\geh=\B.\cr}
$$
The following \VO s are normalised using Dynkin diagram automorphisms.
$$
\eqalign{
\B:&\;
\Phit_{\La_n}^{\La_1 V^{(n)}},%1
\Phit_{\La_1}^{\La_n V^{(n)}},\cr%1
\D:&\;
\Phit_{\La_n}^{\La_1 V^{(n-1)}},%1
\Phit_{\La_{n-1}}^{\La_0 V^{(n-1)}},%2
\Phit_{\La_{n-1}}^{\La_1 V^{(n)}},\cr%12
&~
\Phit_{\La_0}^{\La_n V^{(n-i)}},%3
\Phit_{\La_0}^{\La_{n-1} V^{(n+i-1)}},%31
\Phit_{\La_1}^{\La_n V^{(n+i-1)}},%32
\Phit_{\La_1}^{\La_{n-1}V^{(n-i)}}.\cr}%312
$$
where $i=0$ if $n$ is even and $i=1$ if $n$ is odd.

%*** the following was extensively edited.
%*** also the inequalities were changed in the D case for
%*** conform with the `normal ordering' defined in section 4.

For the fusion representations we normalise as follows:
\par\noindent(i) $\B:$
\par\noindent\quad for $k$: even
$$
\eqalignno{
\Phit_{\La_0}^{\La_0 V^{(k)}}(z)|\La_0\rangle
  &=|\La_0\rangle\ot\hskip-.7cm\sum_{1\le i_1<\cdots<i_m\le n}\hskip-.6cm
  a_{[i_1\cdots i_m\cdots -i_m\cdots -i_1]}
  v_{[i_1\cdots i_m\cdots -i_m\cdots -i_1]}+\cdots,\cr
a_{[i_1\cdots i_m\cdots -i_m\cdots -i_1]}
  &=q^{i_1+\cdots+i_m-(n+1)m}{(1-q^{k})(1-q^{k-2})\cdots(1-q^{k-2m+2})
  \over(1+q)^m},\cr
&\phantom{La}\cr}
$$
\par\noindent\quad for $k$: odd
$$
\eqalignno{
\Phit_{\La_1}^{\La_0 V^{(k)}}(z)|\La_1\rangle
  &=|\La_0\rangle\ot\hskip-.7cm\sum_{1<i_1<\cdots<i_m\le n}\hskip-.6cm
  a_{[1i_1\cdots i_m\cdots -i_m\cdots -i_1]}
  v_{[1i_1\cdots i_m\cdots -i_m\cdots -i_1]}+\cdots,\cr
a_{[1i_1\cdots i_m\cdots -i_m\cdots -i_1]}
  &=q^{i_1+\cdots+i_m-(n+1)m}{(1-q^{k-1})(1-q^{k-3})\cdots(1-q^{k-2m+1})
  \over(1+q)^m},\cr
&\phantom{La}\cr}
$$
\par\noindent\quad for $k$: all
$$
\Phit_{\La_n}^{\La_n V^{(k)}}(z)|\La_n\rangle
   =|\La_n\rangle\ot \alpha^{-k}v_{[0\cdots0]}+\cdots.
$$
The symbol $\cdots$ between $i_m$ and $-i_m$ stands for either a
sequence of pairs $0,0$, or it may be empty, as in
the `normal order' of generating vectors defined in \ref{nord}.

\noindent(ii) $\D:$
\par\noindent\quad for $k$: even
$$
\eqalignno{
\Phit_{\La_0}^{\La_0 V^{(k)}}(z)|\La_0\rangle
  &=|\La_0\rangle\ot\hskip-.7cm\sum_{1\le i_1<\cdots<i_m\le n}\hskip-.6cm
  a_{[i_1\cdots i_m\cdots -i_m\cdots -i_1]}
  v_{[i_1\cdots i_m\cdots -i_m\cdots -i_1]}+\cdots,\cr
a_{[i_1\cdots i_m\cdots -i_m\cdots -i_1]}
  &=q^{i_1+\cdots+i_m-nm}(1-q^{k})(1-q^{k-2})\cdots(1-q^{k-2m+2}),\cr
&\phantom{La}\cr}
$$
\par\noindent\quad for $k$: odd
$$
\eqalignno{
\Phit_{\La_1}^{\La_0 V^{(k)}}(z)|\La_1\rangle
  &=|\La_0\rangle\ot\hskip-.7cm\sum_{1<i_1<\cdots<i_m\le n}\hskip-.6cm
  a_{[1i_1\cdots i_m\cdots -i_m\cdots -i_1]}
  v_{[1i_1\cdots i_m\cdots -i_m\cdots -i_1]}+\cdots,\cr
a_{[1i_1\cdots i_m\cdots -i_m\cdots -i_1]}
  &=q^{i_1+\cdots+i_m-nm}(1-q^{k-1})(1-q^{k-3})\cdots(1-q^{k-2m+1}).\cr}
$$
The symbol $\cdots$ between $i_m$ and $-i_m$ stands for either a
sequence of pairs $-n,n$, or it may be empty, as in
the `normal order' of generating vectors.
The following \VO s are normalised using Dynkin diagram automorphisms.
$$
\eqalignno{
\B:&\;
\Phit_{\La_1}^{\La_1 V^{(2k)}},%1
\Phit_{\La_0}^{\La_1 V^{(2k-1)}},\cr%1
\D:&\;
\Phit_{\La_1}^{\La_1 V^{(2k)}},%1
\Phit_{\La_0}^{\La_1 V^{(2k-1)}},%1
\Phit_{\La_n}^{\La_n V^{(2k)}},%3
\Phit_{\La_{n-1}}^{\La_n V^{(2k-1)}},%3
\Phit_{\La_{n-1}}^{\La_{n-1} V^{(2k)}},%13
\Phit_{\La_n}^{\La_{n-1}V^{(2k-1)}}.\cr}%13
$$

The type II \VO s are normalised in the same manner.
%
%
%               Section 6
%
%
\section{Calculation of two point functions}

In this section we calculate all the two point functions of the form \ref{tpf}
required for computation of the excitation spectrum.
The simplification that one of the \VO s is always for the vector
representation is due to the fact that our underlying physical model is based
on
$\Rb^{(1,1)}(z)$.

\subsection{Dynkin diagram symmetry of two point functions}
Let $\la,\mu$ be level 1 dominant integral weights, let $v$ be a
weight vector in the fundamental representation $V^{(k)}$. We
define a non negative integer $m(\la,\mu;v)$ as the minimal
value of $m_0$ satisfying
$$
\eqalignno{
\la-\mu+&\sum_{j=0}^n m_j\alpha_j \equiv \wt v \quad \mod \Z\delta,\cr
&m_j\ge0\quad(j=0,1,\cdots,n).\cr}
$$
Suppose we have a two point function of the following form:
$$
\eqalignno{
\br{\Phi_\mu^{\nu (V^{(k')})_2}(z_2)\Phi_\la^{\mu (V^{(k)})_1}(z_1)}
&=z_1^{\Delta_\mu-\Delta_\la}z_2^{\Delta_\nu-\Delta_\mu}
\sum_i a_i(z_1/z_2)v_i\ot v'_i.\cr}
$$
Let $\sigma$ be a Dynkin diagram automorphism. Since the \VO s are
normalised using Dynkin diagram automorphisms, we have
$$
\eqalignno{
\br{\Phi_{\sigma(\mu)}^{\sigma(\nu)\sigma(V^{(k')})_2}(z_2)
\Phi_{\sigma(\la)}^{\sigma(\mu)\sigma(V^{(k)})_1}(z_1)}
&=z_1^{\Delta_{\sigma(\mu)}-\Delta_{\sigma(\la)}}
z_2^{\Delta_{\sigma(\nu)}-\Delta_{\sigma(\mu)}}\cr
&\quad\times\sum_i a_i(z_1/z_2)(z_1/z_2)^{m_i}\sigma(v_i)\ot\sigma(v'_i),\cr}
$$
where $m_i=m(\sigma(\la),\sigma(\mu);\sigma(v_i))-m(\la,\mu;v_i)$.
This symmetry reduces the number of cases of two point functions to
calculate.
In the following subsections, we first list all of the cases which are needed,
and then give the explicit formulae, omitting those which are obtained using
the Dynkin diagram symmetry.

Let us explain how to minimise the calculation of two point functions. We are
to calculate the following four kinds.
$$
\leqalignsp{
&\langle\Phi_\mu^{\nu (V^{(k)})_2}(z_2)
\Phi_\la^{\mu (V^{(1)})_1}(z_1)\rangle, &\hbox{(i)}\cr
&\langle\Phi_{\mu'}^{\nu (V^{(1)})_2}(z_2)
\Phi_\la^{\mu' (V^{(k)})_1}(z_1)\rangle, &\hbox{(ii)}\cr
&\langle\Phi_\mu^{(V^{(k)})_2\nu}(z_2)
\Phi_\la^{\mu (V^{(1)})_1}(z_1)\rangle, &\hbox{(iii)}\cr
&\langle\Phi_{\mu'}^{\nu (V^{(1)})_2}(z_2)
\Phi_\la^{\mu' (V^{(k)})_1}(z_1)\rangle, &\hbox{(iv)}\cr}
$$
for $k=1,\cdots,n$. We start with (i). Using \refprop{usefulprop1} (i),
we can obtain the two point function up to scalar function of $z_1,z_2$.
Apply \refprop{qKZ} (a) to determine the scalar part. Calculating (ii) is
similar, but \refprop{usefulprop2} helps us determine the scalar function.
For (iii) and (iv) we use \refprop{usefulprop1} (ii).

\subsection{Spin representations}
The following combinations of weights occur:
$$
\okeqalignno{
(\la,\mu,\mu',\nu)&=(\La_n,\La_n,\La_1,\La_0)
             &k=&n &\hbox{for }\geh= \B, &(spcase:a)\cr
             &=(\La_n,\La_n,\La_0,\La_1)
             &&n &\hbox{for }\geh= \B, &(spcase:b)\cr
             &=(\La_0,\La_1,\La_n,\La_n)
             &&n &\hbox{for }\geh= \B, &(spcase:c)\cr
             &=(\La_1,\La_0,\La_n,\La_n)
             &&n &\hbox{for }\geh= \B, &(spcase:d)\cr
             &=(\La_{n-1},\La_n,\La_1,\La_0)
             &&n &\hbox{for }\geh= \D, &(spcase:e)\cr
             &=(\La_n,\La_{n-1},\La_1,\La_0)
             &&n-1 &\hbox{for }\geh= \D, &(spcase:f)\cr
             &=(\La_{n-1},\La_n,\La_0,\La_1)
             &&n-1 &\hbox{for }\geh= \D, &(spcase:g)\cr
             &=(\La_n,\La_{n-1},\La_0,\La_1)
             &&n &\hbox{for }\geh= \D, &(spcase:h)\cr
             &=(\La_0,\La_1,\La_n,\La_{n-1})
             &&n-i &\hbox{for }\geh= \D, &(spcase:i)\cr
             &=(\La_1,\La_0,\La_n,\La_{n-1})
             &&n+i-1 &\hbox{for }\geh= \D, &(spcase:j)\cr
             &=(\La_0,\La_1,\La_{n-1},\La_n)
             &&n+i-1 &\hbox{for }\geh= \D, &(spcase:k)\cr
             &=(\La_1,\La_0,\La_{n-1},\La_n)
             &&n-i &\hbox{for }\geh= \D, &(spcase:l)\cr}
$$
where $i=0$ if $n$ is even and $i=1$ if $n$ is odd. Due to the Dynkin
diagram symmetry it suffices to give the formulae for \ref{spcase:a},
\ref{spcase:c} and \ref{spcase:e}.

$$
\leqalignsp{
&\langle\Phi_\mu^{\nu (V^{(k)})_2}(z_2)
\Phi_\la^{\mu (V^{(1)})_1}(z_1)\rangle
=z_1^{\Delta_\mu-\Delta_\la}z_2^{\Delta_\nu-\Delta_\mu}
  {(-sq\xi^{5/2}z_1/z_2;\xi^2)_\infty\over(-sq\xi^{3/2}z_1/z_2;\xi^2)_\infty}
  w(z_1/z_2),
&\hbox{(i)}\cr}
$$
where $w(z)$ reads as follows:
$$
\eqalignsp{
w(z)
&=\alpha^{-1}v_0\ot v_{\vep^+}-q^{1/2}\sum_{j=1}^n(-q)^{n-j}v_j\ot
  v_{\vep^+_{(j)}}
  &\hbox{for \ref{spcase:a}},\hskip.5cm\cr
&=\alpha^{-1}sq^n v_0\ot v_{\vep^-}+s\sum_{j=1}^n q^{j-1}v_{-j}\ot
  v_{\vep^{-(j)}}
  &\hbox{for \ref{spcase:c}},\hskip.5cm\cr
&=v_{-n}\ot v_{\vep^+}+\sum_{j=1}^{n-1}(-q)^{n-j}v_j\ot v_{\vep^+_{(j)(n)}}
  &\hbox{for \ref{spcase:e}}.\hskip.5cm\cr}
$$
$$
\leqalignsp{
&\langle\Phi_{\mu'}^{\nu (V^{(1)})_2}(z_2)
  \Phi_\la^{\mu'(V^{(k)})_1}(z_1)\rangle
=z_1^{\Delta_{\mu'}-\Delta_\la}z_2^{\Delta_\nu-\Delta_{\mu'}}
  {(-sq\xi^{5/2}z_1/z_2;\xi^2)_\infty\over(-sq\xi^{3/2}z_1/z_2;\xi^2)_\infty}
  w(z_1/z_2),
  &\hbox{(ii)}\cr}
$$
where $w(z)$ reads as follows:
$$
\eqalignsp{
w(z)
&=\alpha^{-1}(-q)^n v_{\vep^+}\ot v_0+\sum_{j=1}^n(-q)^{j-1}v_{\vep^+_{(j)}}
  \ot v_j&\hbox{for \ref{spcase:a}},\hskip.5cm\cr
&=\alpha^{-1}v_{\vep^-}\ot v_0+\sum_{j=1}^n q^{n-j+1/2}v_{\vep^{-(j)}}
  \ot v_{-j}&\hbox{for \ref{spcase:c}},\hskip.5cm\cr
&=(-q)^{n-1}v_{\vep^+}\ot v_{-n}+\sum_{j=1}^{n-1}(-q)^{j-1}v_{\vep^+_{(j)(n)}}
  \ot v_j&\hbox{for \ref{spcase:e}}.\hskip.5cm\cr}
$$
$$
\leqalignsp{
&\langle \Phi_\mu^{(V^{(k)})_2 \nu}(z_2)
\Phi_\la^{\mu(V^{(1)})_1}(z_1)\rangle
=z_1^{\Delta_\mu-\Delta_\la}z_2^{\Delta_\nu-\Delta_\mu}
  {(-s\xi^{5/2}z_1/z_2;\xi^2)_\infty\over(-s\xi^{3/2}z_1/z_2;\xi^2)_\infty}
  w(z_1/z_2),
&\hbox{(iii)}\cr}
$$
where $w(z)$ reads as follows:
$$
\eqalignsp{
w(z)
&=\alpha^{-1}v_0\ot v_{\vep^+}-q^{1/2}\sum_{j=1}^n(-q)^{n-j}v_j\ot
v_{\vep^+_{(j)}}
  &\hbox{for \ref{spcase:a}},\hskip.5cm\cr
&=\alpha^{-1}sq^{n-1/2}v_0\ot v_{\vep^-}+s\sum_{j=1}^n q^{j-1}v_{-j}\ot
  v_{\vep^{-(j)}}
  &\hbox{for \ref{spcase:c}},\hskip.5cm\cr
&=v_{-n}\ot v_{\vep^+}+\sum_{j=1}^{n-1}(-q)^{n-j}v_j\ot v_{\vep^+_{(j)(n)}}
  &\hbox{for \ref{spcase:e}}.\hskip.5cm\cr}
$$
$$
\leqalignsp{
&\langle \Phi_{\mu'}^{\nu (V^{(1)})_2}(z_2)
\Phi_\la^{(V^{(k)})_1\mu'}(z_1)\rangle
=z_1^{\Delta_{\mu'}-\Delta_\la}z_2^{\Delta_\nu-\Delta_{\mu'}}
{(-s\xi^{1/2}z_1/z_2;\xi^2)_\infty\over(-s\xi^{-1/2}z_1/z_2;\xi^2)_\infty}
w(z_1/z_2),&\hbox{(iv)}\cr}
$$
where $w(z)$ reads as follows:
$$
\eqalignsp{
w(z)
&=\alpha^{-1}q^{1/2}(-q)^{-n}v_{\vep^+}\ot v_0+\sum_{j=1}^n(-q)^{1-j}
  v_{\vep^+_{(j)}}\ot v_j
  &\hbox{for \ref{spcase:a}},\hskip.5cm\cr
&=\alpha^{-1}v_{\vep^-}\ot v_0+\sum_{j=1}^n q^{j-n-1/2}v_{\vep^{-(j)}}\ot
v_{-j}
  &\hbox{for \ref{spcase:c}},\hskip.5cm\cr
&=(-q)^{1-n}v_{\vep^+}\ot v_{-n}+\sum_{j=1}^{n-1}(-q)^{1-j}v_{\vep^+_{(j)(n)}}
  \ot v_j
  &\hbox{for \ref{spcase:e}}.\hskip.5cm\cr}
$$

\subsection{Vector and fusion representations}
Note that $1\le k\le n-1$ for $\B$ ($n-2$ for $\D$) in this subsection.
The following combinations of weights occur:
$$
\okeqalignno{
(\la,\mu,\mu',\nu)&=(\La_1,\La_0,\La_1,\La_0)
             &k:&\hbox{even} &\hbox{for }\geh= \B,\D, &(fuscase:a)\cr
             &=(\La_0,\La_1,\La_1,\La_0)
             &&\hbox{odd} &\hbox{for }\geh= \B,\D, &(fuscase:b)\cr
             &=(\La_1,\La_0,\La_0,\La_1)
             &&\hbox{odd} &\hbox{for }\geh= \B,\D, &(fuscase:c)\cr
             &=(\La_0,\La_1,\La_0,\La_1)
             &&\hbox{even} &\hbox{for }\geh= \B,\D, &(fuscase:d)\cr
             &=(\La_n,\La_n,\La_n,\La_n)
             &&\hbox{all} &\hbox{for }\geh= \B, &(fuscase:e)\cr
             &=(\La_{n-1},\La_n,\La_{n-1},\La_n)
             &&\hbox{even} &\hbox{for }\geh= \D, &(fuscase:f)\cr
             &=(\La_n,\La_{n-1},\La_{n-1},\La_n)
             &&\hbox{odd} &\hbox{for }\geh= \D, &(fuscase:g)\cr
             &=(\La_{n-1},\La_n,\La_n,\La_{n-1})
             &&\hbox{odd} &\hbox{for }\geh= \D, &(fuscase:h)\cr
             &=(\La_n,\La_{n-1},\La_n,\La_{n-1})
             &&\hbox{even} &\hbox{for }\geh= \D. &(fuscase:i)\cr}
$$
Due to the Dynkin diagram symmetry it suffices to give the formulae for
\ref{fuscase:a},\ref{fuscase:b} and \ref{fuscase:e}.
$$
\leqalignsp{
&\langle
\Phi_\mu^{\nu (V^{(k)})_2}(z_2)\Phi_\la^{\mu (V^{(1)})_1}(z_1)
\rangle
&\hbox{(i)}\cr
&\>=z_1^{\Delta_\mu-\Delta_\la}z_2^{\Delta_\nu-\Delta_\mu}
{((-q)^{k+1}\xi^2z_1/z_2;\xi^2)_\infty((-q)^{-k+1}\xi^3z_1/z_2;\xi^2)_\infty
\over((-q)^{k+1}\xi
z_1/z_2;\xi^2)_\infty((-q)^{-k+1}\xi^2z_1/z_2;\xi^2)_\infty}
w^{(i)}(z_1/z_2),}
$$
where $w^{(i)}(z)$ reads as follows:

\noindent
For $\B$:
$$
\eqalignsp{
w^{(1)}(z)
&=\sum c_{[i_1\cdots i_m]}v_1\ot
  v_{[i_1\cdots i_m \cdots -i_m\cdots -i_1]}\cr
&\quad+\sum_{j\neq1} c_{-j[i_1\cdots i_m]}v_j\ot
  v_{[1i_1\cdots i_m \cdots -i_m\cdots -j\cdots -i_1]}\cr
&\quad+\sum_{j\neq1} c_{j[i_1\cdots i_m]}v_{-j}\ot
  v_{[1i_1\cdots j\cdots i_m \cdots -i_m\cdots -i_1]}\cr
&\quad+\sum c_{0[i_1\cdots i_m]}v_0\ot
  v_{[1i_1\cdots i_m \cdots -i_m\cdots -i_1]}
                   &\hbox{for \ref{fuscase:a}},\hskip.5cm\cr
w^{(2)}(z) &=\sum c'_{-j[i_1\cdots i_m]}v_j\ot
  v_{[i_1\cdots i_m \cdots -i_m\cdots -j\cdots -i_1]}\cr
&\quad+\sum c'_{j[i_1\cdots i_m]}v_{-j}\ot
  v_{[i_1\cdots j\cdots i_m \cdots -i_m\cdots -i_1]}\cr
&\quad+\sum c'_{0[i_1\cdots i_m]}v_0\ot
  v_{[i_1\cdots i_m \cdots -i_m\cdots -i_1]}
                   &\hbox{for \ref{fuscase:b}},\hskip.5cm\cr
w^{(3)}(z)
&=\alpha^{-(k+1)}(1-(-q)^k)\cr
&\qquad\times\sum_{j=1}^n(q^{n-j}v_j\ot v_{[0\cdots 0-j]}
  +q^{n+j-2}z v_{-j}\ot v_{[j0\cdots 0]})\cr
&\quad+\alpha^{-(k+1)}(1+q^{2n}z)v_0\ot v_{[0\cdots0]}
  &\hbox{for \ref{fuscase:e}}.\hskip.5cm\cr}
$$
The symbol $\cdots$ between $i_m$ and $-i_m$ stands for either a sequence
of pairs $0,0$, or it may be empty.
The coefficients in the foregoing formulae are given by
$$
\eqalignsp{
c_{[i_1\cdots i_m]}
&=q^{i_1+\cdots+i_m-(n+1)m}\cr
&\times{(1-q^k)(1-q^{k-2})\cdots(1-q^{k-2m+2})\over(1+q)^m}
  \cases{(1+q^{2n-k}z)&\quad($i_1\neq1$)\cr (1-q^{2n}z)&\quad($i_1=1$),\cr}\cr
\rcases{c_{-j[i_1\cdots i_m]}\cr c_{j[i_1\cdots i_m]}\cr}
&=(-q)^{l+1} q^{i_1+\cdots+i_m-(n+1)m}
  &(i_l<j<i_{l+1})\hskip.5cm\cr
  &\qquad\times{(1-q^k)(1-q^{k-2})\cdots(1-q^{k-2m})
  \over(1+q)^{m+1}}(1+q^{-k})z
\cases{q^{n-j}\cr q^{j-n-1},\cr}\cr
c_{0[i_1\cdots i_m]}
&=(-q)^{m+1} q^{i_1+\cdots+i_m-(n+1)m}\cr
  &\qquad\times{(1-q^k)(1-q^{k-2})\cdots(1-q^{k-2m})
  \over(1+q)^{m+1}}(1+q^{-k})z,\cr
\rcases{c'_{-j[i_1\cdots i_m]}\cr c'_{j[i_1\cdots i_m]}\cr}
&=s(-q)^l q^{i_1+\cdots+i_m-(n+1)m}
  &(i_l<j<i_{l+1})\hskip.5cm\cr
  &\qquad\times{(1-q^{k-1})(1-q^{k-3})\cdots(1-q^{k-2m+1})
  \over(1+q)^m}\cases{q^{2n-j}\cr q^{j-1},\cr}\cr
c'_{0[i_1\cdots i_m]}
&=s(-q)^m q^{i_1+\cdots+i_m-(n+1)m}\cr
  &\qquad\times{(1-q^{k-1})(1-q^{k-3})\cdots(1-q^{k-2m+1})
  \over(1+q)^m}q^n.\cr}
$$
For $\D$:
$$
\eqalignsp{
w^{(1)}(z)
&=\sum c_{[i_1\cdots i_m]}v_1\ot
  v_{[i_1\cdots i_m \cdots -i_m\cdots -i_1]}\cr
&\quad+\sum_{j\neq1} c_{-j[i_1\cdots i_m]}v_j\ot
  v_{[1i_1\cdots i_m \cdots -i_m\cdots -j\cdots -i_1]}\cr
&\quad+\sum_{j\neq1} c_{j[i_1\cdots i_m]}v_{-j}\ot
  v_{[1i_1\cdots j\cdots i_m \cdots -i_m\cdots -i_1]}
                   &\hbox{for \ref{fuscase:a}},\hskip.5cm\cr
w^{(2)}(z) &=\sum c'_{-j[i_1\cdots i_m]}v_j\ot
  v_{[i_1\cdots i_m \cdots -i_m\cdots -j\cdots -i_1]}\cr
&\quad+\sum c'_{j[i_1\cdots i_m]}v_{-j}\ot
  v_{[i_1\cdots j\cdots i_m \cdots -i_m\cdots -i_1]}
                   &\hbox{for \ref{fuscase:b}}.\hskip.5cm\cr}
$$
The symbol $\cdots$ between $i_m$ and $-i_m$ stands for either a sequence
of pairs $-n,n$, or it may be empty.
The coefficients in the foregoing formulae are given by
$$
\eqalignsp{
c_{[i_1\cdots i_m]}
&=q^{i_1+\cdots+i_m-nm}\cr
&\times(1-q^k)(1-q^{k-2})\cdots(1-q^{k-2m+2})
  \cases{(1+q^{2n-k-1}z)&\quad($i_1\neq1$)\cr
             (1-q^{2n-1}z)&\quad($i_1=1$),\cr}\cr
\rcases{c_{-j[i_1\cdots i_m]}\cr c_{j[i_1\cdots i_m]}\cr}
&=(-q)^{l+1} q^{i_1+\cdots+i_m-nm}
  &(i_l<j<i_{l+1})\hskip.5cm\cr
  &\qquad\times(1-q^k)(1-q^{k-2})\cdots(1-q^{k-2m})
  (1+q^{-k})z \cases{q^{n-j}\cr q^{j-n},\cr}\cr
\rcases{c'_{-j[i_1\cdots i_m]}\cr c'_{j[i_1\cdots i_m]}\cr}
&=s(-q)^l q^{i_1+\cdots+i_m-nm}
  &(i_l<j<i_{l+1})\hskip.5cm\cr
  &\qquad\times(1-q^{k-1})(1-q^{k-3})\cdots(1-q^{k-2m+1})
  \cases{q^{2n-j-1}\cr q^{j-1}.\cr}\cr}
$$
$$
\leqalignsp{
&\langle
\Phi_{\mu'}^{\nu(V^{(1)})_2}(z_2)
\Phi_\la^{\mu'(V^{(k)})_1}(z_1)
\rangle
&\hbox{(ii)}\cr
&=z_1^{\Delta_{\mu'}-\Delta_\la}z_2^{\Delta_\nu-\Delta_{\mu'}}
{((-q)^{k+1}\xi^2z_1/z_2;\xi^2)_\infty((-q)^{-k+1}\xi^3z_1/z_2;\xi^2)_\infty
\over((-q)^{k+1}\xi
z_1/z_2;\xi^2)_\infty((-q)^{-k+1}\xi^2z_1/z_2;\xi^2)_\infty}
\ol{w}^{(i)}(z_1/z_2),}
$$
where $\ol{w}^{(i)}(z)$ reads as follows:
$$
\eqalignsp{
\ol{w}^{(1)}(z)
&=q^2\xi^2zP(q^{-2\ol{\rho}-\ol{\La}_1}\ot 1)w^{(1)}(q^{-2}\xi^{-2}z^{-1})
                   &\hbox{for \ref{fuscase:a}},\hskip.5cm\cr
\ol{w}^{(2)}(z)
&=P(q^{-2\ol{\rho}}\ot 1)w^{(2)}(z)
                   &\hbox{for \ref{fuscase:b}},\hskip.5cm\cr
\ol{w}^{(3)}(z)
&=q\xi zP(q^{-2\ol{\rho}-2\ol{\La}_n}\ot 1)w^{(3)}(q^{-2}\xi^{-2}z^{-1})
                   &\hbox{for \ref{fuscase:e}}.\hskip.5cm\cr}
$$
$$
\leqalignsp{
&\langle \Phi_\mu^{(V^{(k)})_2 \nu}(z_2)
 \Phi_\la^{\mu(V^{(1)})_1}(z_1)\rangle&\hbox{(iii)}\cr
&\>=z_1^{\Delta_\mu-\Delta_\la}z_2^{\Delta_\nu-\Delta_\mu}
  {(-(-q)^k\xi^2z_1/z_2;\xi^2)_\infty(-(-q)^{-k}\xi^3 z_1/z_2;\xi^2)_\infty
 \over(-(-q)^k\xi z_1/z_2;\xi^2)_\infty(-(-q)^{-k}\xi^2z_1/z_2;\xi^2)_\infty}
w(z_1/z_2),\cr}
$$
where $w(z)$ reads as follows:
$$
\eqalignsp{
w(z)&=w^{(1)}(q^{-1}z)&\hbox{for \ref{fuscase:a}},\hskip.5cm\cr
    &=w^{(2)}(z)&\hbox{for \ref{fuscase:b}},\hskip.5cm\cr
    &=(q^{-\ol{\La}_n}\ot 1)w^{(3)}(q^{-1}z)
                &\hbox{for \ref{fuscase:e}}.\hskip.5cm\cr}
$$
$$
\leqalignsp{
&\langle \Phi_{\mu'}^{\nu (V^{(1)})_2}(z_2)
\Phi_\la^{(V^{(k)})_1 \mu'}(z_1)\rangle&\hbox{(iv)}\cr
&\>=z_1^{\Delta_{\mu'}-\Delta_\la}z_2^{\Delta_\nu-\Delta_{\mu'}}
  {(-(-q)^kz_1/z_2;\xi^2)_\infty(-(-q)^{-k}\xi z_1/z_2;\xi^2)_\infty
 \over(-(-q)^k\xi^{-1}z_1/z_2;\xi^2)_\infty(-(-q)^{-k}z_1/z_2;\xi^2)_\infty}
\ol{w}(z_1/z_2),\cr}
$$
where $\ol{w}(z)$ reads as follows:
$$
\eqalignsp{
\ol{w}(z)&=(q^{-2\ol{\rho}-\ol{\La}_1}\ot 1)\ol{w}^{(1)}(q^{-1}\xi^{-2}z)
                &\hbox{for \ref{fuscase:a}},\hskip.5cm\cr
         &=\xi^{-1}(q^{-2\ol{\rho}}\ot 1)\ol{w}^{(2)}(z)
                &\hbox{for \ref{fuscase:b}},\hskip.5cm\cr
         &=(q^{-2\ol{\rho}-\ol{\La}_n}\ot 1)\ol{w}^{(3)}(q^{-1}\xi^{-2}z)
                &\hbox{for \ref{fuscase:e}}.\hskip.5cm\cr}
$$
%
%
%               Section 7
%
%
\section{Commutation relations}

\subsection{\VO s for dual representations}
Here we define and normalise the \VO s of type I
$\Phi_\la^{\mu(V^{(1)})^{*a^{\pm1}}}(z)$
and of type II
$\Phi_\la^{(V^{(k)})^{*a^{-1}}\mu}(z)$.
They are used for the mathematical formulation of the local structure,
transfer matrix and creation operators in 7.3.

For the former case, we recall the isomorphism $C_\pm^{(1)}$
\ref{dual1}, and define
$$
\Phit_\la^{\mu(V^{(1)})^{*a^{\pm1}}}(z)
=(\hbox{const})(\id\ot C_\pm^{(1)})\cdot
\Phit_\la^{\mu V^{(1)}}(z\xi^{\mp1}).
$$
The constant prefactor is so chosen that we have
$$
\Phit_\la^{\mu(V^{(1)})^{*a^{\pm1}}}(z)\ket{\la}
=\ket{\mu}\ot \gamma v_j^*+\cdots,
$$
where $v_j$ is a base vector in $V^{(1)}$ such that $\wt v_j=\mu-\la$,
and $\gamma=1$ except that when $\la=\mu=\La_n$ for
$\B$, then $\gamma= q^{-1/2}\alpha^{-1}$ ($\alpha$ is defined in
\ref{defalpha}).
This choice of $\gamma$ makes \ref{locstr} and
\ref{trvac} hold for any possible pair $(\la,\mu)$.

For the \VO s $\Phi_\la^{(V^{(k)})^{*a^{-1}}\mu}(z)$ of type II, using
\ref{dualsp} and \ref{dualk}, we simply define
$$
\eqalign{
\Phi_\la^{(V^{(k)})^{*a^{-1}}\mu}(z)
&=s(C_-^{(k')}\ot\id)\cdot\Phi_\la^{(V^{(k')})\mu}(z\xi^{-1})
\qquad(k:\hbox{spin and }\La_0\in\{\la,\mu\}),\cr
&=\phantom{s}(C_-^{(k')}\ot\id)\cdot\Phi_\la^{(V^{(k')})\mu}(z\xi^{-1})
\qquad(\hbox{otherwise}).\cr}
$$
Here, if $V^{(k)}$ is for the vector or fusion representation, then we need
$k'=k$, but if it is for the spin representation, then $C_-^{(k')}$ should
be understood as $C_-^{(sp)}$ and $k'$ of $V^{(k')}$ should be chosen
carefully.
(The isomorphism is explained directly after \ref{dualsp}.) $s$ is defined
in \ref{defs}.
This definition is made so that part (3) of \refprop{com} below will hold
for all possible choices of $(\la,\mu,\mu',\nu)$, without any extra
prefactors.

\subsection{Commutation relations}
Using the results of section 6 and definitions of subsections 5.3 and 7.1,
we can prove the following commutation relations for \VO s.

\prop{com}. For any possible combination of weights $(\la,\mu)$ or
$(\la,\mu,\mu',\nu)$, we have
$$
\eqalign{
\hbox{\rm(1)}\qquad&
\Phi_\mu^{\la(V^{(1)})_2}(z_2)\Phi_\la^{\mu(V^{(1)})_1}(z_1)
=r(z_1/z_2)\Rb^{(1,1)}(z_1/z_2)
\Phi_\mu^{\la(V^{(1)})_1}(z_1)\Phi_\la^{\mu(V^{(1)})_2}(z_2),\cr
\hbox{\rm(2)}\qquad&
\Phi_{\mu'}^{\nu(V^{(1)})_2}(z_2)\Phi_\la^{(V^{(k)})_1\mu'}(z_1)
=\tau^{(k)}(z_1/z_2)
\Phi_\mu^{(V^{(k)})_1 \nu}(z_1)\Phi_\la^{\mu(V^{(1)})_2}(z_2),\cr
\hbox{\rm(3)}\qquad&
\Phi_{\mu'}^{\nu(V^{(1)})_2}(z_2)\Phi_\la^{(V^{(k)})^{*a^{-1}}_1\mu'}(z_1)
=\tau^{(k)}(z_1/z_2)^{-1}
\Phi_\mu^{(V^{(k)})^{*a^{-1}}_1\nu}(z_1)\Phi_\la^{\mu(V^{(1)})_2}(z_2).\cr}
$$
Here
$$
\eqalignsp{
r(z)&=z^{-1}
{(q^2z;\xi^2)_\infty(\xi z;\xi^2)_\infty
(q^2\xi z^{-1};\xi^2)_\infty(\xi^2z^{-1};\xi^2)_\infty \over
(q^2\xi z;\xi^2)_\infty(\xi^2z;\xi^2)_\infty
(q^2z^{-1};\xi^2)_\infty(\xi z^{-1};\xi^2)_\infty},\cr
\tau^{(k)}(z)&=z^{-1}{\Th_{\xi^2}(-(-q)^k z)\Th_{\xi^2}(-(-q)^{-k}\xi z)
       \over\Th_{\xi^2}(-(-q)^k z^{-1})\Th_{\xi^2}(-(-q)^{-k}\xi z^{-1})},
       &\hbox{\rm(fusion)}\hskip.5cm\cr
       &=z^{-1/2}{\Th_{\xi^2}(-s\xi^{1/2}z)
       \over\Th_{\xi^2}(-s\xi^{1/2}z^{-1})}.
       &\hbox{\rm(spin)}\hskip.5cm\cr}
$$
\endprop
We only need to show them at the level of vacuum expectation value (two point
function). See proposition 6.1 of \cite{DFJMN92}.

\subsection{Mathematical formulations}
In this subsection, we briefly review the symmetry approach \cite{DFJMN92,
IIJMNT92}, and collect formulae needed for the validity of this approach.
$V$ is again to be understood as $V^{(1)}$.

In section 2 we have defined the space of states to be
$$
\eqalign{
\F&=\bigoplus_{i,j}\F_{\La_i\La_j},\cr
\F_{\la\mu}&=V(\la)\ot V(\mu)^{*a}\simeq{\Hom}_{\Q(q)}(V(\mu),V(\la)),\cr}
$$
where $i,j$ run over $0,1,n$ (also $n-1$ for $\D$). The vacuum (ground state)
vector $\vac_\la\in\F_{\la\la}$ is defined as $\id_{V(\la)}$. Following
\cite{IIJMNT92}, we define the left action of $\uq$ on $\F_{\la\mu}$. We
can also define the right action on the same underlying space. This right
module is denoted by $\F^r_{\la\mu}$.
There is a natural pairing:
$$
\langle f \mid g \rangle=
{\tr_{V(\la)}(q^{-2\rho}fg) \over \tr_{V(\la)}(q^{-2\rho})},
\qquad f\in\F^r_{\la\mu},~g\in\F_{\mu\la}.
\eqno(pairing)
$$

Next we show that $\Phit_\la^{\mu V}(z)$ gives an isomorphism from $V(\la)$
to $V(\mu)\ot V_z$. Define
$$
p:\quad V^{*a}\ot V \longrightarrow  \Q(q), \qquad
        v_1^*\ot v_2 \mapsto \langle v_1^*,v_2 \rangle.
$$
Then the above statement is a consequence of the following formulae. (See
proposition 4.1 of \cite{IIJMNT92}.)
$$
\eqalignno{
p\left(\langle \Phit_\mu^{\la V^{*a}_1}(z)\Phit_\la^{\mu V_2}(z) \rangle\right)
&=g,&(locstr:a)\cr
\langle \Phit_\mu^{\la V_1}(z)\Phit_\la^{\mu V^{*a}_2}(z) \rangle
&=g\sum_{i\in J}v_i\ot v_i^*.&(locstr:b)\cr}
$$
Here
$$
g={(q^2\xi;\xi^2)_\infty(\xi^2;\xi^2)_\infty \over
   (q^2;\xi^2)_\infty(\xi;\xi^2)_\infty}.\eqno(defg)
$$
Iterating this $\Phit_\la^{\mu V}(z)$ we see the local structure of our
space of states.

We proceed to the mathematical formulation to the row transfer matrix.
Recall the definitions of $\la^{(k)},\la^{(k)*}$ in subsection 2.5 and the
discussion in subsection 1.2.
The row transfer matrix
$$
T(z)=T_{\la\mu}^{\la^{(1)}\mu^{(1)}}(z) :\quad
\F_{\la\mu}\longrightarrow\F_{\la^{(1)}\mu^{(1)}}
$$
is formulated as the composition of the following operators:
$$
V(\la)\ot V(\mu)^{*a}\longrightarrow
V(\la^{(1)})\ot V_z\ot V(\mu)^{*a}\longrightarrow
V(\la^{(1)})\ot V(\mu^{(1)})^{*a}.
$$
Here the first and the second maps are given by
$\Phit_\la^{\la^{(1)}V}(z)\ot\id$ and
$\id\ot\left(\Phit_{\mu^{(1)}}^{\mu V^{*a^{-1}}}(z)\right)^t$.
We can show
$$
T_{\la\la}^{\la^{(1)}\la^{(1)}}(z)\vac_\la=g\vac_{\la^{(1)}},
$$
from the formula (See 4.3 of \cite{IIJMNT92}.):
$$
\ol{p}\left(
\langle \Phit_\mu^{\la V_1}(z)\Phit_\la^{\mu V^{*a^{-1}}_2}(z) \rangle\right)
=g,\eqno(trvac)
$$
where $\ol{p}$ is defined by
$$
\ol{p}:\quad V\ot V^{*a^{-1}} \longrightarrow  \Q(q), \qquad
        v_1\ot v_2^* \mapsto \langle v_1,v_2^* \rangle.
$$

Finally we recall the formulation of creation and annihilation operators.
Let $I$ be an index of base vectors in $V^{(k)}$. Decompose the following
type II \VO s into components:
$$
\eqalign{
\Phi_\la^{V^{(k)}\la^{(k)}}(z)
&=\sum_I v_I\ot \Phi_{\la,I}^{(k)}(z),\cr
\Phi_\la^{(V^{(k)})^{*a^{-1}}\la^{(k)*}}(z)
&=\sum_I v_I^*\ot \Phi_{\la,I}^{(k)*}(z).\cr}
$$
The creation operator $\phi_{\la,I}^{(k)*}(z)$ is defined by
$$
\phi_{\la,I}^{(k)*}(z):\quad
\F_{\la\mu}\longrightarrow\F_{\la^{(k)*}\mu},\qquad
f\mapsto\Phi_{\la,I}^{(k)*}(z)\circ f.
$$
The annihilation operator $\phi_{\la,I}^{(k)}(z)$ is defined by the adjoint
of
$$
\F^r_{\mu\la^{(k)}}\longrightarrow\F^r_{\mu\la},\qquad
f\mapsto f\circ\Phi_{\la,I}^{(k)}(z),
$$
with respect to the pairing \ref{pairing}. In both cases the quasi-momentum
$z$ is supposed to be on $|z|=1$.

In conclusion, the commutation relations between the transfer matrix and the
creation  (annihilation) operators given in section 2 are direct consequences
of
parts (2) and (3) of \refprop{com}.

\numberby{}
\prefixby{A}
%
%
%               Appendix A
%
%
\section*{Appendix A}

\par\noindent{\sl A.1 Proof of \refprop{defW}.\hskip1em}
We shall describe $\Im R(q^2)$ rather than $W$ (see \ref{Rvec} and
\ref{rdefR} for the definition of $R(z)$).
Since $R$--matrices preserve the weight, we can concentrate on each weight
space in $V\ot V$.
For the weight spaces of non-zero weight $2\eta_1\ep_l$ and
$\eta_1\ep_l+\eta_2\ep_m$ ($\eta_1,\eta_2=\pm1,l,m=1,\cdots,n$), it is easy to
see that they are generated by the vectors $v_i\ot v_i$ and
$qv_i\ot v_j+q^{2\theta(j\prec i)}v_j\ot v_i$ ($i=\eta_1 l,j=\eta_2 m$).

The weight space of weight 0 is generated by the following
vectors:
$$
A_j=\sum_{i\in J}\bar{a}_{ij}v_i\ot v_{-i}\qquad(j\in J),
$$
where $\bar{a}_{ij}=a_{ij}(q^2)/(1-q^2)$, and is given explicitly by
$$
\eqalignno{
\bar{a}_{ij}
&=\cases{q^2(1-\xi)&($i=j\ne0$)\cr
         q^2(1-\xi)+q(1-q^2\xi)&($i=j=0$)\cr
         q^{\ol{j}-\ol{i}}(q^2-1)+\delta_{i,-j}(1-q^2\xi)
               &($i\prec j$)\cr
         q^2(\xi q^{\ol{j}-\ol{i}}(q^2-1)+\delta_{i,-j}(1-q^2\xi))
               &($i\succ j$).\cr}
\cr}
$$
For convenience we put $u_j=v_j\ot v_{-j}$ ($j\in J$). In what follows
in the proof we assume $1\le i,j\le n$, and for $\geh=\D$ we ignore
the term containing $u_0$ and also $A_0$. Noting $\xi=q^{N-2}$ and
recalling the definition \ref{jbar} we get
$$
q^{-j}A_j-q^jA_{-j}=(1-q^N)\{(q^2-1)\sum_{i<j}(q^{-i}u_i+q^iu_{-i})
+(q^{2-j}-q^j)(u_j+u_{-j})\}.
$$
Setting
$$
B_j=(q^2-1)\sum_{i<j}(q^{-i}u_i+q^iu_{-i})+(q^{2-j}-q^j)(u_j+u_{-j}),
$$
we obtain
$$
B_{j+1}-B_j=(q^j-q^{-j})(u_j+q^2u_{-j}-q(u_{j+1}+u_{-j-1})).
$$
Therefore we have
$$
u_j+q^2u_{-j}-q(u_{j+1}+u_{-j-1})\in \Im R(q^2)\qquad(j=1,\cdots,n-1).
$$
%***
Let $U$ be the subspace of $\Im R(q^2)$ generated by these vectors.
Using $(u_j+q^2u_{-j})\equiv q(u_{j+1}+u_{-j-1})~(\mod U)$ we can show
$$
\left.\eqalign{
q^{-2j}A_j&\equiv
\cases{q^{n-j}(1-q)(u_n+q^2u_{-n}-q(1+q)u_0)&for $\B$\cr
       0&for $\D$\cr}\cr
A_0&\equiv-(1-q^{2n})(u_n+q^2u_{-n}-q(1+q)u_0)\cr}\right\}~(\mod U).
$$
This completes the proof.
\qed

\smallskip\par\noindent{\sl A.2 Proof of \ref{rk}.\hskip1em}
We define $R^{\br{k}}$ by
$$
R^{\br{k}}=\lim_{q\to1}{R(q^{-2k})\over1-q^2}.
$$
For the definition of $R(z)$ see \ref{Rvec} and \ref{rdefR}. Then
from \ref{defT} $T^{\br{k}}$ is obtained by
$$
\eqalignno{
T^{\br{k}}&=\widehat{S}_{1k}\cdots\widehat{S}_{13}\widehat{S}_{12},\cr
\widehat{S}_{ij}&=R^{\br{1}}_{j-1j}R^{\br{2}}_{j-2j}\cdots
R^{\br{j-i}}_{ij}.\cr}
$$
Calculating explicitly we have
$$
\eqalignno{
R^{\br{k}}&=P-kI+{2k\over \hc-2k}\sum_{i,j\in J}E_{ij}\ot E_{-i-j}.
&(limR)\cr}
$$

Let $l$ be an integer such that $0\le l\le k$, $k-l:\hbox{even}$. Let
$v_{[i_1\cdots i_k]}$ stand for $v_{i_1}\ot\cdots\ot v_{i_k}$ in
$V^{\ot k}$ as opposed to section 4. We define for $i_1,\cdots,i_l
\in J$
$$
w^{(k)}_{[i_1\cdots i_l]}=
{\sum_{i_{l+1},\cdots, i_k\in J}}^{\hskip-14pt(1)}\hskip7pt
{\sum_{\tau\in S_k}}^{\hskip-2pt(2)}\mathop{\rm sgn}\tau\cdot
v_{[i_{\tau(1)}\cdots i_{\tau(k)}]}.
$$
Here $S_k$ denotes the $k$--th symmetric group, and the two summations
are restricted as
$$
\eqalignno{
{\sum}^{(1)}:~&i_{l+1}+i_{l+2}=0,~i_{l+3}+i_{l+4}=0,~\cdots~,~
              i_{k-1}+i_k=0,\cr
{\sum}^{(2)}:~&\tau^{-1}(l+1)<\tau^{-1}(l+3)<\cdots\tau^{-1}(k-1),\cr
             &\tau^{-1}(l+1)<\tau^{-1}(l+2),~
              \tau^{-1}(l+3)<\tau^{-1}(l+4),\cr
             &\hskip3.5cm\cdots\quad\tau^{-1}(k-1)<\tau^{-1}(k).\cr}
$$
It is easy to see that $w^{(k)}_{[i_1\cdots i_l]}$ has the following
properties:
$$
\eqalignno{
w^{(k)}_{[i_{\tau(1)}\cdots i_{\tau(l)}]}
&=\mathop{\rm sgn}\tau\cdot w^{(k)}_{[i_1\cdots i_l]}\qquad\hbox{for }
\tau\in S_l,&(propw:a)\cr
w^{(k)}_{[i_1\cdots i_l]}
&=\sum_{1\le m\le l}(-)^{m-1}v_{i_m}\ot
w^{(k-1)}_{[i_1\cdots\widehat{i_m}\cdots i_l]}
+(-)^l\sum_{p\in J}v_p\ot w^{(k-1)}_{[i_1\cdots i_l-p]},&(propw:b)\cr}
$$
where $\widehat{~~}$ means omission. Now we prepare

\lem{lem1}.
$$
\eqalignno{
\widehat{S}&_{1k+1}(w^{(k)}_{[i_1\cdots i_l]}\ot v_j)\cr
&=(-)^kk!{\hc-k-l\over \hc-2k}w^{(k+1)}_{[i_1\cdots i_lj]}
+k!{k-l+2\over \hc-2k}\sum_{1\le m\le l}(-)^{m-1}\delta_{i_m,-j}
w^{(k+1)}_{[i_1\cdots\widehat{i_m}\cdots i_l]}.\cr}
$$
\endlem

\Proof
First, from \ref{propw:b} and \ref{limR} we have
$$
\eqalignno{
R^{\br{k}}_{1k+1}&(w^{(k)}_{[i_1\cdots i_l]}\ot v_j)\cr
&=\sum_{1\le m\le l}(-)^{m-1}\Bigl(
v_j\ot w^{(k-1)}_{[i_1\cdots\widehat{i_m}\cdots i_l]}\ot v_{i_m}
-kv_{i_m}\ot w^{(k-1)}_{[i_1\cdots\widehat{i_m}\cdots i_l]}\ot v_j\cr
&\hskip3cm
+{2k\over\hc-2k}\delta_{i_m,-j}\sum_{p\in J}
v_p\ot w^{(k-1)}_{[i_1\cdots\widehat{i_m}\cdots i_l]}\ot v_{-p}
\Bigr)\cr
&\quad+(-)^l\sum_{p\in J}\Bigl(
v_j\ot w^{(k-1)}_{[i_1\cdots i_l-p]}\ot v_p
-kv_p\ot w^{(k-1)}_{[i_1\cdots i_l-p]}\ot v_j\cr
&\hskip3cm
+{2k\over\hc-2k}v_p\ot w^{(k-1)}_{[i_1\cdots i_lj]}\ot v_{-p}
\Bigr).\cr}
$$
We prove the lemma by induction on $k$. Note that
$\widehat{S}_{1k+1}=\widehat{S}_{2k+1}R^{\br{k}}_{1k+1}$, and assume
the formula for $k-1$. After some calculation using the properties of
$w^{(k)}_{[i_1\cdots i_l]}$, we get the desired formula.
\qed

To complete the proof, it suffices to show

\lem{lem2}.
\item{(i)} $w^{(k)}_{[i_1\cdots i_l]}\in\Im T^{\br{k}}$ for all
$i_1,\cdots,i_l\in J$,
\item{(ii)} The vectors $\{w^{(k)}_{[i_1\cdots i_l]}\mid
l=k,k-2,\cdots,0\hbox{ or }1,i_1\prec i_2\prec\cdots\prec i_l\}$
are linearly independent.
\endlem

\Proof
We prove (i) by the induction on $k$. From the induction assumption
we know that $w^{(k-1)}_{[i_1\cdots i_{l-1}]}\in \Im T^{\br{k-1}}$.
 From \ref{propw:a} we can assume $i_1,\cdots,i_{l-1}$ are distinct.
We divide the proof into the following cases:
\par\noindent(a)\quad
$i_m+i_l\ne0$ for $m=1,\cdots,l-1$,
\par\noindent(b)\quad
there exists $a$ ($1\le a\le l-1$) such that $i_a+i_l=0$,
\par\noindent(c)\quad$l=0$.

\noindent
Put
$$
A=(-)^{k-1}(k-1)!{\hc-k-l+2\over\hc-2k+2},\qquad
B=(k-1)!{k-l+2\over\hc-2k+2}.
$$
For (a) see \reflem{lem1}. We have
$$
\widehat{S}_{1k}(w^{(k-1)}_{[i_1\cdots i_{l-1}]}\ot v_{i_l})
=Aw^{(k)}_{[i_1\cdots i_l]}\in\widehat{S}_{1k}(\Im T^{\br{k-1}}\ot V)
=\Im T^{\br{k}}.
$$
For (b) we similarly have
$$
\eqalignno{
\widehat{S}_{1k}(w^{(k-1)}_{[i_1\cdots i_a\cdots i_{l-1}]}\ot v_{-i_a})
&=Aw^{(k)}_{[i_1\cdots i_a\cdots i_{l-1}-i_a]}
+Bw^{(k)}_{[i_1\cdots\widehat{i_a}\cdots i_l]},\cr
\widehat{S}_{1k}(w^{(k-1)}_{[i_1\cdots-i_a\cdots i_{l-1}]}\ot v_{i_a})
&=Aw^{(k)}_{[i_1\cdots-i_a\cdots i_{l-1}i_a]}
+Bw^{(k)}_{[i_1\cdots\widehat{-i_a}\cdots i_l]}.\cr}
$$
Recalling that $w^{(k)}_{[i_1\cdots i_a\cdots i_{l-1}-i_a]}=
-w^{(k)}_{[i_1\cdots-i_a\cdots i_{l-1}i_a]}$, we get
$$
w^{(k)}_{[i_1\cdots i_a\cdots i_{l-1}-i_a]},
w^{(k)}_{[i_1\cdots\widehat{i_a}\cdots i_{l-1}]}\in\Im T^{\br{k}}.
$$
%***
For (c) apply (b) to the case $l=2$.

We proceed to the proof of (ii).
The proof is reduced to showing
linear independence of the vectors with the same weight.
So we
are to show for $\alpha_1,\cdots,\alpha_l\in J$ ($\alpha_1\prec
\cdots\prec\alpha_l$)
$$
\eqalignno{
{\sum_I}^*a_I&
w^{(k)}_{[\alpha_1\cdots\alpha_li_1\cdots i_m-i_m\cdots-i_1]}=0
\Longrightarrow a_I=0.&(star)\cr}
$$
Here $\sum^*$ is the summation on $I=(i_1\cdots i_m)$ such that
$1\le i_1<\cdots<i_m\le n,i_p\ne\alpha_q$ for any $(p,q)$.
We use induction on $k$.
Since the case $l=k$ is trivial, we assume $l<k$.
Note the following simple fact:
$$
\sum_{j\in J}v_j\ot w_j=0\hbox{ for }w_j\in V^{\ot(k-1)}
\Longrightarrow w_j=0\hbox{ for all }j.
$$
Fix $j>0$ such that $j\ne\pm\alpha_1,\cdots,\pm\alpha_l$.
Look at the first component of the sum \ref{star} and use the above
observation, then as coefficients of $v_j$ and $v_{-j}$ we have
$$
\eqalignno{
0&=\sum_I\theta(i_1=j)a_{[ji_2\cdots i_m]}(-)^l
w^{(k-1)}_{[\alpha_1\cdots\alpha_li_2\cdots i_m-i_m\cdots -i_2-j]}\cr
&+\sum_I\theta(i_2=j)a_{[i_1ji_3\cdots i_m]}(-)^{l+1}
w^{(k-1)}_{[\alpha_1\cdots\alpha_li_1i_3\cdots-i_3-j-i_1]}\cr
&\hskip3cm\vdots\cr
&=\sum_I\theta(i_m=j)a_{[i_1\cdots i_{m-1}j]}(-)^{l+m-1}
w^{(k-1)}_{[\alpha_1\cdots\alpha_li_1\cdots i_{m-1}-j-i_{m-1}\cdots -i_1]}\cr
&+\theta(m\ne{k-l\over2})\sum_Ia_I(-)^l
w^{(k-1)}_{[\alpha_1\cdots\alpha_li_1\cdots i_m-i_m\cdots -i_1-j]},\cr
0&=\sum_I\theta(i_1=j)a_{[ji_2\cdots i_m]}(-)^{l+1}
w^{(k-1)}_{[\alpha_1\cdots\alpha_lji_2\cdots i_m-i_m\cdots -i_2]}\cr
&+\sum_I\theta(i_2=j)a_{[i_1ji_3\cdots i_m]}(-)^{l+2}
w^{(k-1)}_{[\alpha_1\cdots\alpha_li_1ji_3\cdots-i_3-i_1]}\cr
&\hskip3cm\vdots\cr
&=\sum_I\theta(i_m=j)a_{[i_1\cdots i_{m-1}j]}(-)^{l+m}
w^{(k-1)}_{[\alpha_1\cdots\alpha_li_1\cdots i_{m-1}j-i_{m-1}\cdots -i_1]}\cr
&+\theta(m\ne{k-l\over2})\sum_Ia_I(-)^l
w^{(k-1)}_{[\alpha_1\cdots\alpha_li_1\cdots i_m-i_m\cdots -i_1j]}\cr}
$$
{}From the induction assumption we have
$$
a_{[i_1\cdots i_p j i_{p+1}\cdots i_m]}\pm a_{[i_1\cdots i_m]}=0
\quad\hbox{for all }I=(i_1\cdots i_m),
$$
where $p$ is given from $i_p<j<i_{p+1}$. This completes the proof.
\qed

\smallskip\par\noindent{\sl A.3 Proof of \ref{dualk}.\hskip1em}
In this subsection we use the following notations.
$$
\eqalignno{
&\Delta^{(k)\prime}(x)=\sum x_{(k)}\ot\cdots\ot x_{(1)}\quad\hbox{if }
\Delta^{(k)}(x)=\sum x_{(1)}\ot\cdots\ot x_{(k)},\cr
&P^{(k)}\hbox{ is the linear operator }
P^{(k)}v_1\ot\cdots\ot v_k=v_k\ot\cdots\ot v_1\,(v_j\in V),\cr
&\Tc^{(k)}=P^{(k)}T^{(k)}.\cr}
$$
We prepare two lemmas.
The first can be shown directly.

\lem{copant}.
$$
\eqalignno{
&\Delta^{(k)}\circ a^{\pm1}
=(a^{\pm1}\ot\cdots\ot a^{\pm1})\circ\Delta^{(k)\prime},\cr
&\Delta^{(k)\prime}\circ a^{\pm1}
=(a^{\pm1}\ot\cdots\ot a^{\pm1})\circ\Delta^{(k)}.\cr}
$$
\endlem

\lem{d.s.}.
\item{(1)} $\Tc^{(k)}$ is an isomorphism on $\Im \Tc^{(k)}$.
\item{(2)} $V^{\ot k}=\Im \Tc^{(k)}\oplus \Ker T^{(k)}$.
\endlem

\Proof
Since $\Im\Tc^{(k)}\cong V^{(k)}$ as a $\uq$--module, we know from
\ref{decomp} that
$$
\eqalignno{
\Im\Tc^{(k)}&\cong V_{\ol{\La}_k}\oplus V_{\ol{\La}_{k-2}}\oplus\cdots\oplus
V_{\ol{\La}_1}\hbox{ or }V_0 \quad\hbox{as a $U_q(\ol{{\geh}})$--module}.\cr}
$$
Therefore we have
$$
\eqalignno{
\Tc^{(k)}
&=\la_k{\id}_{V_{\ol{\La}_k}}+\la_{k-2}{\id}_{V_{\ol{\La}_{k-2}}}+\cdots
\cr}
$$
with some scalars $\la_k,\la_{k-2},\cdots\neq0$. For (2) it suffices to
show that $\Im\Tc^{(k)}\cap\Ker T^{(k)}=0$. Taking $\Tc^{(k)}v\in
\Im\Tc^{(k)}\cap\Ker T^{(k)}$, we have $(\Tc^{(k)})^2v=0$. From (1) we get
$\Tc^{(k)}v=0$.
\qed

In what follows, we explicitly write the $\uq$--module structure in such a
manner as $(V_{z(-q)^{1-k}}\ot\cdots\ot V_{z(-q)^{k-1}})^{\ast a^{\pm1}}$,
$V_{z(-q)^{1-k}}\ot'\cdots\ot' V_{z(-q)^{k-1}}$, etc. Here $\ot'$ signifies
the use of the opposite coproduct $\Delta'$. Let us start by noting an
isomorphism
$$
\eqalignno{
V^{(k)}_z&{\buildrel\rm def\over =}
(V_{z(-q)^{1-k}}\ot\cdots\ot V_{z(-q)^{k-1}})\big/\Ker T^{(k)}\cr
&\iso T^{(k)}(V_{z(-q)^{1-k}}\ot\cdots\ot V_{z(-q)^{k-1}}).\cr}
$$
Consider the following sequence of homomorphisms:
$$
\eqalignno{
(V^{(k)}_z)^{\ast a^{\pm1}}
&=\bigl((V_{z(-q)^{1-k}}\ot\cdots\ot V_{z(-q)^{k-1}})
\big/\Ker T^{(k)}\bigr)^{\ast a^{\pm1}}\cr
%*** I assume that K is \Q(q)
&\iso\{f:V_{z(-q)^{1-k}}\ot\cdots\ot V_{z(-q)^{k-1}}\rightarrow\Q(q)\mid
%*** I changed the notation f| here
f:\Ker T^{(k)}\rightarrow0\}&(mod)\cr
&\hookrightarrow
(V_{z(-q)^{1-k}}\ot\cdots\ot V_{z(-q)^{k-1}})^{\ast a^{\pm1}}\cr
&\iso(V_{z(-q)^{1-k}})^{\ast a^{\pm1}}\ot'\cdots\ot'
(V_{z(-q)^{k-1}})^{\ast a^{\pm1}}&(iso1)\cr
&\iso V_{z\xi^{\mp1}(-q)^{1-k}}\ot'\cdots\ot' V_{z\xi^{\mp1}(-q)^{k-1}}.
&(iso2)\cr}
$$
The isomorphism \ref{iso1} is due to \refLem{copant}, and \ref{iso2} is
given by $C^{-1}_\pm\ot\cdots\ot C^{-1}_\pm$, where $C_\pm=C^{(1)}_\pm$
is defined in \ref{dual1}. The remaining thing is to show that the image
of \ref{mod} in \ref{iso2} is identified as
$T^{(k)}(V_{z\xi^{\mp1}(-q)^{1-k}}\ot\cdots\ot V_{z\xi^{\mp1}(-q)^{k-1}})$.
First we can show
$$
\eqalignno{
\{f&:V_{z(-q)^{1-k}}\ot\cdots\ot V_{z(-q)^{k-1}}\rightarrow\Q(q)\mid
f:\Ker T^{(k)}\rightarrow0\}\cr
&\hskip1.5cm\iso T^{(k)\ast}(V_{z(-q)^{1-k}}\ot'\cdots\ot'
V_{z(-q)^{k-1}})^{\ast a^{\pm1}},\cr
f&\mapsto \Tc^{(k)\ast}f,\cr}
$$
in the following manner. We can easily show that the map is an homomorphism.
Noting \refLem{d.s.}(2), for $g:V_{z(-q)^{1-k}}\ot'\cdots\ot'V_{z(-q)^{k-1}}
\rightarrow \Q(q)$ we set
$$
\eqalignno{
f(v)&=\cases{g(P^{(k)}v)&$v\in\Im \Tc^{(k)}$\cr
             0&$v\in\Ker T^{(k)}$.\cr}\cr}
$$
We can show $\Tc^{(k)\ast}f=T^{(k)\ast}g$. Thus we have shown the surjectivity.
The injectivity can also be shown using \refLem{d.s.}(2).

To finish we have to show that the image of $T^{(k)\ast}(V_{z(-q)^{1-k}}\ot'
\cdots\ot'V_{z(-q)^{k-1}})^{\ast a^{\pm1}}$ by $C_\pm^{-1}\ot\cdots\ot
C_\pm^{-1}$ is $T^{(k)}(V_{z\xi^{\mp1}(-q)^{1-k}}\ot\cdots\ot
V_{z\xi^{\mp1}(-q)^{k-1}})$, or
$$
\eqalignno{
&(C_\pm^{-1}\ot\cdots\ot C_\pm^{-1})T^{(k)t}(C_\pm\ot\cdots\ot C_\pm)
=T^{(k)}.\cr}
$$
This reduces to checking
$(C_\pm^{-1}\ot C_\pm^{-1})\Rb(z)^t(C_\pm\ot C_\pm)=\Rb(z)$, which can be
done by a direct calculation.

\section*{Acknowledgements}

We would like to thank
%***
M. Batchelor,
E. Date,
M. Jimbo,
A. Kuniba,
T. Nakanishi,
N. Yu. Reshetikhin,
V. Rittenberg,
H. Yamane
%***
for helpful discussions.
Special thanks are due to J. Suzuki for sending a fax of his
Bethe Ansatz calculation.
This work was partly supported by grants from the ARC (SARC6600410),
the SMS and Grant-in-Aid for Scientific Research on Priority Areas,
the Ministry of Education, Science and Culture, Japan.
M. Okado would like to thank the Mathematics Department of the ANU for
hospitality during two visits to Canberra, when much of the work was done.

\section*{References}

\def\PT{Physics Today}

\refis{F89} \jnlitem
{Feynman's Office: The last blackboards}{\PT}{}{February, (1989), p88}

\refis{ABF84} \jnlitem
{G. E. Andrews, R. J. Baxter and P. J. Forrester:
Eight-vertex SOS model and generalized Rogers-Ramanujan type identities}
{\JSP}{35}{(1984), 193-266}

\refis{Bax76} \jnlitem
{R. J. Baxter: Corner transfer matrices of the eight vertex model.
Low temperature expansions and conjectured properties}
{\JSP}{15}{(1976), 485-503}

\refis{Bax77} \jnlitem
{R. J. Baxter: Corner transfer matrices of the eight vertex model.
II. The Ising Model Case}
{\JSP}{17}{(1977), 1-14}
%Corner Transfer Matrices

\refis{Bax80} \jnlitem
{R. J. Baxter:
Hard Hexagons: exact solution}
{\JPA}{13}{(1980), L61-L70}
\jnlitem{--- :
Rogers-Ramanujan identities in the hard hexagon model}
{\JSP}{26}{(1981), 427-452}

\refis{BDV83} \jnlitem
{O. Babelon, H. J. de Vega and C. M. Viallet,
Exact excitation spectra of the $\Z_{n+1}\times\Z_{n+1}$
generalized Heisenberg model}
{\NPB}{220}{[FS8] (1983), 283--301}

\refis{Bou} \bkitem
{N. Bourbaki}
{Groupes et alg\`ebres de Lie, \'El\'ements de math\`ematique}
{Hermann, Paris 1968}

\refis{Dav93} \jnlitem
{B. Davies:
Corner transfer matrices and quantum affine algebras}
{\JPA}{27}{(1994), 361-378}

\refis{Dav94} \jnlitem
{B. Davies:
Infinite dimensional symmetry of corner transfer matrices}
{\IJMPA}{}{to appear}

\refis{DFJMN92} \jnlitem
{B. Davies, O. Foda, M. Jimbo, T. Miwa and A. Nakayash\-iki:
Diagonalization of the XXZ Hamiltonian by vertex operators}
{\CMP}{151}{(1993), 89-153}

\refis{DJKMO87} \jnlitem
{E. Date, M. Jimbo, A. Kuniba, T. Miwa and M. Okado:
Exactly solvable SOS models: Local height probabilities and
theta function identities}
{\NPB}{290}{(1987), 231-273}

\refis{DJO93} \jnlitem
{E. Date, M. Jimbo and M. Okado:
Crystal base and $q$-vertex operators}
{\CMP}{155}{(1993), 47-69}

\refis{DO93} \jnlitem
{E. Date and M. Okado: Calculation of excitation spectra of the
spin model related with the vector representation of the
quantized aff\-ine algebra of type $A^{(1)}_n$}
{\IJMPA}{9}{(1994), 399-417}
%higher rank case

\refis{FM92} \jnlitem
{O. Foda and T. Miwa:
Corner transfer matrices and quantum affine algebras}
{\IJMPA}{7}{{\it Suppl.\/} 1A, (1992), 279-302}

\refis{FR92} \jnlitem
{I. B. Frenkel and N. Yu. Reshetikhin:
Quantum affine algebras and holonomic difference equations}
{\CMP}{149}{(1992), 1-60}

\refis{IIJMNT92} \jnlitem
{M. Idzumi, K. Iohara, M. Jimbo, T. Miwa, T. Nakashima and
T. Tokihiro: Quantum affine symmetry in vertex models}
{\IJMPA}{8}{(1993), 1479-1511}

\refis{J86} \jnlitem
{M. Jimbo,
Quantum $R$ matrix for the generalized Toda system}
{\CMP}{102}{(1986), 537--547}

\refis{JMO88b} \jnlitem
{M. Jimbo, T. Miwa and M. Okado:
Local state probabilities of solvable lattice models:
An $A^{(1)}_{n-1}$ family}
{\NPB}{300}{(1988), 74-108}
%LSP for affine models

\refis{JMO92} \jnlitem
{M. Jimbo, T. Miwa and Y. Ohta:
Structure of the space of states in RSOS models}
{\IJMPA}{8}{(1993), 1457-1477}

\refis{JMMN92} \jnlitem
{M. Jimbo, K. Miki, T. Miwa and A. Nakaya\-shiki:
Correlation Functions of the \XXZ\ mod\-el for $\Delta<-1$}
{\PLA}{168}{(1992), 256-263}

\refis{JMN92} \jnlitem
{M. Jimbo, T. Miwa and A.  Nakayashiki:
Difference equations for the correlation functions
of the eight-vertex model}
{\JPA}{26}{(1993), 2199-2209}

\refis{Kac90} \bkitem
{V. G. Kac}
{Infinite dimensional Lie algebras, 3rd ed.}
{Cambridge University Press, Cambridge 1990}

\refis{ORW87} \jnlitem
{E. Ogievetsky, N. Yu. Reshetikhin and P. Wiegmann:
The principal chiral field in two dimensions on classical Lie algebras}
{\NPB}{18}{(1987), 45-98}

\refis{SW83} \jnlitem
{K. Sogo and M. Wadati:
Boost operator and its application to quantum Gelfand-Levitan
equation for Heisenberg-Ising chain with spin one-half}
{\goodbreak\PTP}{69}{(1983), 431-450}

\refis{Th86} \jnlitem
{H. B. Thacker:
Corner transfer matrices and Lorentz invariance on a lattice}
{\PD}{18}{(1986), 348-359}
%lorenz invariance on lattice

\refis{Th86} \jnlitem
{H. B. Thacker:
Corner transfer matrices and Lorentz invariance on a lattice}
{\PD}{18}{(1986), 348-359}

\refis{KNS94} \jnlitem
{A. Kuniba, T. Nakanishi and J. Suzuki:
Functional relations in solvable lattice models II: Applications}
{\IJMPA}{9}{(1994), 5267-5312}

\refis{Ko} \jnlitem
{N. Jing, S.-J. Kang and Y. Koyama:
Vertex operators of quantum affine Lie algebras $U_q(\D)$}
{\CMP}{}{to appear}

\refis{Na} \jnlitem
{T. Nakanishi:
Fusion, mass, and representation theory of the Yangian algebra}
{\NPB}{439}{(1995), 441-460}

\listreferences

\bye